\documentclass[aps,pra,twocolumn,amsfonts,amssymb,amsmath,10pt,showpacs,showkeys,superscriptaddress]{revtex4-1}
\usepackage{graphicx}
\usepackage{dcolumn}%
\usepackage{bm}%

\begin{document}
\title{Spin relaxation in a one-dimensional large-spin degenerate Fermi gas}
\date{\today}

 \author{Ulrich~Ebling}
  \email[Electronic address:]{ebling@cat.phys.s.u-tokyo.ac.jp} 
 \affiliation{Department of Physics, University of Tokyo, 7-3-1 Hongo, Bunkyo-ku, Tokyo 113-0033, Japan} 
 \affiliation{Max-Planck-Institut f\"{u}r Physik komplexer Systeme, N\"othnitzer Str. 38, D-01187 Dresden, Germany}

 \author{Andr\'e~Eckardt}
 \affiliation{Max-Planck-Institut f\"{u}r Physik komplexer Systeme, N\"othnitzer Str. 38, D-01187 Dresden, Germany}

\begin{abstract}
In this work, we study the dynamics of an atomic harmonically trapped large-spin Fermi gas in one dimension (1D). We investigate the interplay of different collision processes.  Coherent spin oscillations, driven by spin-changing forward scattering are captured by a mean-field description and scale linearly with density regardless of the dimension of the system. Conversely, ``incoherent`` collision processes which e.g.\ lead to the damping of spin oscillations, behave differently. In the usual three-dimensional (3D) case, the rate of incoherent processes increases faster with density than mean-field effects, but in 1D it increases slower. This means, that in the 1D case, incoherent collisions become more important at lower densities. We study these effects by deriving and integrating a quantum Boltzmann equation. We demonstrate that the well known fact that in one dimension, interaction-induced correlations become more dominant at low densities can be observed in far-from-equilibrium spin dynamics.
\end{abstract}

\maketitle

\section{Introduction}
\begin{figure}[t]
\centering
\includegraphics[width=8.6cm]{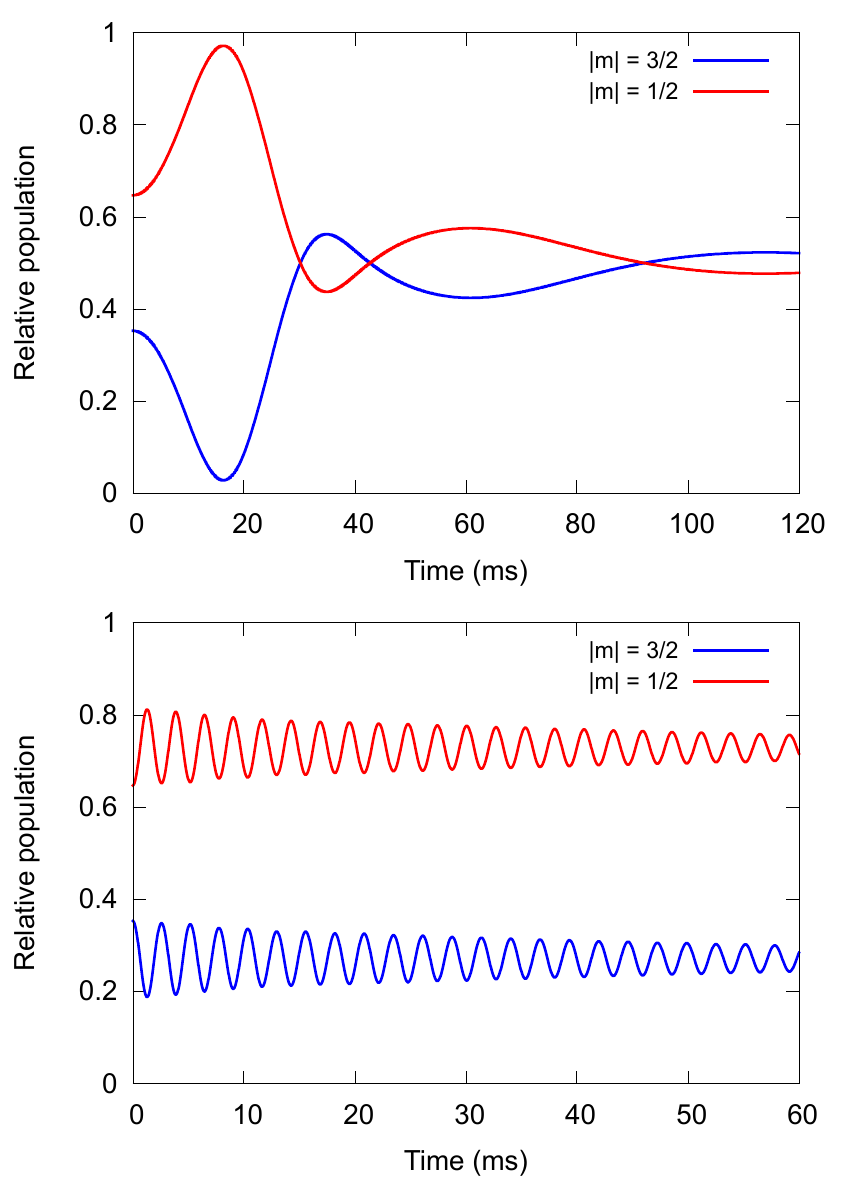}
\caption{Spin oscillations of a spin 3/2 Fermi gas for weak and strong magnetic fields. Top: $B=0.1\:\text{G}$, the dynamics is dominated by the non-linear mean-field term in Eq.~(\ref{eq:9}). Bottom: $B=1\:\text{G}$, the dynamics is dominated by the quadratic Zeeman effect and features damped harmonic oscillations with a single frequency and damping rate. In the remainder of this paper, we consider this regime when extracting frequencies and damping rates.}
\label{fig1}
\end{figure}

Ultracold atomic gases have distinguished themselves by offering physicists access to clean, isolated quantum many-body systems with a large degree of control. This makes them ideal systems to study coherent many-body dynamics and to address the question, whether and how an isolated quantum many-body system thermalizes. Another feature of trapped quantum gases is the ability to reduce the dimension of the gas by applying a strong confinement along one or two axes such that arrays of 2D ``disks`` or 1D ``tubes`` \cite{Goerlitz2001,Greiner2001} can be created, which are decoupled and can be described individually. This makes it possible to study lower-dimensional systems, which are expected to show different behavior from regular 1D systems, especially regarding interactions and correlations \cite{Tolra2004,Fallani2014}, with prominent examples the appearance of a confinement-induced resonance \cite{Olshanii1998} or the realization of a Tonks-Girardeau gas \cite{Kinoshita2004,Paredes2004}. In addition, precise experimental control over magnetic fields and preparation of complex spin configurations has promoted theoretical and experimental studies of spinor dynamics and related properties of large-spin fermions \cite{Ho1999,Wu2003,Lecheminant2005,Cazalilla2009,Gorshkov2010,Krauser2012,Krauser2014,Ebling2014,Ho2015} and spin diffusion effects in ultracold spinor gases \cite{McGuirk2002,Nikuni2002,Fuchs2003,Endo2008,Du2008,Natu2009,Piechon2009,Ebling2011,Heinze2013,Koschorrek2013,Enss2015,Trotzky2015,Koller2016}.

In this paper we investigate spinor dynamics in a dimensionally reduced system, an ultracold 1D Fermi gas with large spin. We focus on multicomponent systems where interactions are not $SU(N)$-symmetric because of differences in s-wave scattering lengths for different total spin scattering channels \cite{Krauser2012}, which leads to the presence of spin-changing collisions. Further, we consider a weakly-interacting gas in an intermediate regime between the hydrodynamic and collisionless cases. In such a system, the leading interaction effect is two-body forward scattering, where not only total kinetic energy is conserved, but also the individual momenta of each atom involved in a scattering event. Such interactions can be treated on a mean-field level, as they appear as coherent processes on the single-particle level and therefore, the effect of forward scattering is a phase shift. Also present in our considerations are collisions induced by backward scattering, where the individual momenta of particles are changed so that these collisions appear as dissipative processes on the single-particle level.

Dynamics induced by forward scattering can take the form of spin rotations or oscillating population dynamics in homogeneous magnetic fields \cite{Pechkis2013,Krauser2014} or spin waves and transport in the presence of inhomogeneous magnetic fields \cite{McGuirk2002,Piechon2009,Heinze2013,Koller2016} and can be described with a mean field approach that results in a collisionless Boltzmann equation \cite{Piechon2009,Natu2009,Ebling2011}. 
For such effects, dimensionality only plays a role by rescaling the scattering lengths with a factor depending on the ratio of trapping frequencies, without a qualitative change in behavior under parameter change. What we denote by collisions are scattering events in which individual momenta of particles change. In contrast to the forward scattering mentioned above, such collisions appear as incoherent on the single particle level and can be taken into account by adding a collision integral to the Boltzmann equation. For spin dynamics, collisions lead to damping of spin waves, damping of coherent mean-field oscillations as well as long-time redistribution among spin states \cite{Ebling2014}. These effects are expected to behave differently for different dimension of the  system. It is a well-known fact that in 1D systems, correlations are enhanced at low densities \cite{Giamarchi}. Similarly, in this paper, we observe a relative increase of coherent processes compared to incoherent ones when the density is increased. Here, the collision rate has a sub-linear growth with density and therefore grows slower with increasing density than mean-field interactions. This behavior contrasts the one observed in 3D, where the growth of the collision rate grows faster with the density than mean-field interactions, up to the point that for very large densities, collisions entirely block mean-field dynamics and stabilize the system in an originally unstable spin configuration \cite{Krauser2014}.

In this paper, we investigate the interplay between mean-field dynamics and ``collisions`` for a range of experimentally realizable parameters. As shown in Fig.~\ref{fig1}, we investigate coherent spin oscillations in a 1D gas, in a regime where the amplitude is relatively small and hence the oscillations are approximately harmonic with exponential damping. The oscillation frequency provides us with a measure for the strength of coherent mean-field interactions, while the damping rate gives us a measure of the collision rate. Our theoretical description is in terms of a quantum Boltzmann equation for the single-particle Wigner function of a multicomponent 1D Fermi gas in a harmonic trap and homogenous magnetic field.

\section{System}
In experiments, 1D quantum gases are created by applying a 2D optical lattice to a sample of atoms trapped e.g. in an optical dipole trap. This optical lattice can be tuned to be so strong that it splits the atom cloud into 1D tubes that are not coupled, and the transversal confinement can be considered harmonic with a frequency $\omega_\perp$. If the condition $\omega_\perp > N\omega_x$ is satisfied, for a sufficiently cold gas with temperature $k_B T\ll\hbar\omega_\perp$ we can assume all particles to occupy the transversal ground state, so that we can treat the system as a Fermi gas in a one-dimensional harmonic trap of frequency $\omega\equiv\omega_x$. For an arbitrary spin $F$ and quadratic Zeeman splitting $Q$, we describe such a system with the Hamiltonian
\begin{align}
\label{eq:1}
\hat H=&\int dx\sum_m \hat\psi_m^\dagger(x)\left[\frac{\hbar^2\nabla^2}{2M}+\frac12M\omega^2 x^2+Qm^2\right]\hat\psi_m(x)\nonumber\\
&+\frac12\int dx\sum_{klmn}U_{klmn}\hat\psi_k^\dagger(x)\hat\psi_m^\dagger(x)\hat\psi_n(x)\hat\psi_l(x).
\end{align}
The coupling constants $U_{klmn}$ in the interaction part of the Hamiltonian are given by
\begin{equation}
\label{eq:2}
 U_{klmn}=\sum_{S,M}g_S\langle km|SM\rangle\langle SM|ln\rangle,
\end{equation}
where $\langle km|SM\rangle$ denote the Clebsch Gordan coefficients for a pair of spins with individual magnetic quantum numbers $k$ and $m$ to form total spin $\lbrace S,M\rbrace$, and $g_S=2\hbar\omega_\perp a_S$ is proportional to the s-wave scattering length for this total spin $S$ channel. The pre-factor $2\hbar\omega_\perp$ appears due to the transversal confinement \cite{Olshanii1998}.

\section{Kinetic Theory}
We describe the time evolution of the system in terms of the single-particle Wigner function
\begin{equation}
\label{eq:3}
W_{mn}(x,p)=\int dy\langle\hat\psi_m^\dagger(x+\tfrac y 2)\hat\psi_n(x-\tfrac y 2)\rangle, 
\end{equation}
the phase-space representation of the single-particle density matrix. The Boltzmann equation for the time-evolution of $W_{mn}(x,p)$ is given by
\begin{align}
\label{eq:4}
\frac{d}{dt} W(x,p)+\partial_0 W(x,p)+\frac{i}{\hbar}\left[Q S_z^2+V^\text{mf}(x), W(x,p)\right]\nonumber\\
-\frac12\left\lbrace\partial_x V^\text{mf}(x),\partial_p W(x,p)\right\rbrace=C[W(x,p)]. 
\end{align}
It consists of a term $\partial_0=p\partial_x/M-M\omega^2 x\partial_p$ that describes the free motion in the trap, a commutator $[,]$ governing coherent spin dynamics induced by the quadratic Zeeman effect and the mean-field potential $V_\text{mf}$, an anticommutator $\lbrace,\rbrace$ describing spin-dependent forces, as well as a collision integral $C$.

As we present in more detail in the appendix, the interaction terms in Eq.~({\ref{eq:4}) are obtained by truncating the Bogoliubov-Born-Green-Kirkwood-Yvon (BBGKY) hierarchy for the many-body problem at the two-body level and closing it by finding a suitable approximation for the two-body density matrix in terms of single-body density matrices, as well as performing a semi-classical gradient expansion. The mean-field terms are originally the leading term of the collision integral, however it is convenient to separate them and add them to the free motion side of the equation. The mean-field potential 
\begin{equation}
\label{eq:5}
V_{mn}^\text{mf}(x)=\int dp \sum_{kl} U_{klnm} W_{kl}(x,p) 
\end{equation}
acts as a density-dependent correction to the trapping potential and the magnetic field. The mean-field terms arise from the real part of the T-matrix for low energy two-body scattering, while the origins of the remaining collision integral are the imaginary part of the T-matrix and its square
\cite{Jeon1989}.

\begin{figure}[t]
\centering
\includegraphics[width=8.6cm]{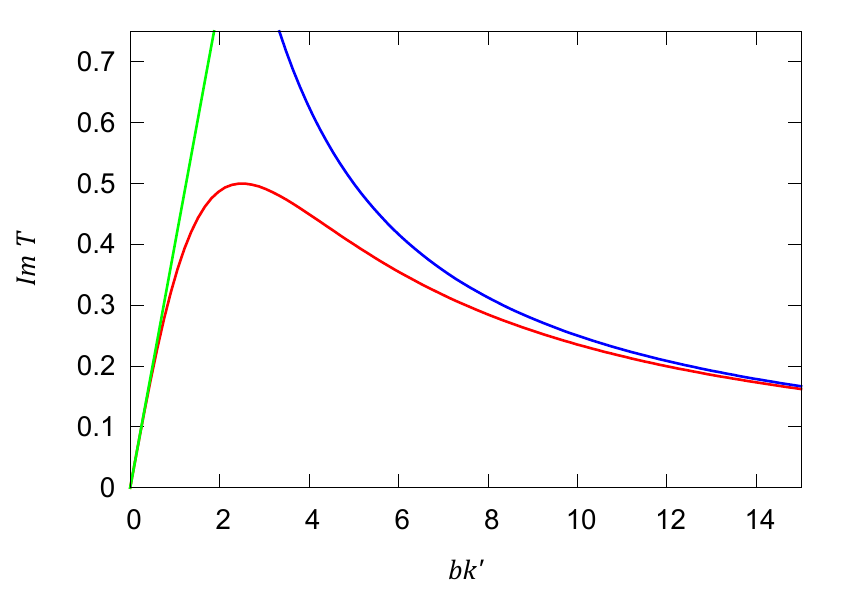}
\caption{Dependence of the imaginary part of the on-shell T-matrix Eq.~(\ref{eq:6}) on the outgoing wave vector $k'$ in units given by the trapping frequency $\omega$ (red line) over a range typical for the systems we consider in this paper. For low s-wave scattering length $a$, $b\to 0$ in 3D (green) and $b\to \infty$ (blue) in 1D, the asymptotic behavior differs. Consequently, the non-forward scattering rate determined by the imaginary part of the T-matrix has a different density behavior depending on the dimension of the system.}
\label{fig2}
\end{figure}

The T-matrix for a two-body collision at low energies, $T(\vec k',\vec k)=\langle \vec k'|T|\vec k\rangle$, is defined as the amplitude for two scattering particles to transition from a state with wave vectors $\lbrace\vec k_1,\vec k_2\rbrace$ into one with $\lbrace\vec k_3,\vec k_4\rbrace$ and in our case only depends on the relative wave vectors $\vec k=\vec k_1-\vec k_2$ and $\vec k'=k_3-k_4$. It has a different dependence on the scattering length for different dimensionalities. If for the moment we assume spin-independent scattering with a single s-wave scattering length $a$, the on-shell T-matrix has the form
\begin{equation}
\label{eq:6}
 T(\vec k,\vec k')=A\frac{b}{1-i|\vec k'|b}.
\end{equation}
for both the 1D and the 3D case. It has the same dependence on the length of the outgoing relative wave vector $\vec k'$, related to the incoming wave vector by conservation of total kinetic and quadratic Zeeman energy (on-shell condition), but the other parameters $A$ and $B$ differ between the 1D and 3D case. In 3D, $A=(2\pi)^{-3}4\pi\hbar^2/M,\,b=a$ and in 1D, $A=(2\pi)^{-1}2\hbar^2 M,\,b=\omega_\perp a/\hbar M$. The 1D scattering length is  inversely proportional to the 3D scattering length of the atoms confined in the one-dimensional tube: $a_{1D}=\hbar(M\omega_\perp a)^{-1}$. The different dependence on the scattering length for both expressions means that in the weakly-interacting limit, the asymptotic behavior is 
\begin{equation}
\label{eq:7}
T(\vec k,\vec k')=\frac{1}{(2\pi)^3}\frac{4\pi\hbar^2a}{M}\left(1+ik'a+\ldots\right)
\end{equation}
 in 3D and 
\begin{equation}
\label{eq:8}
T(k,k')=\frac{1}{2\pi}2\hbar\omega_\perp a\left(1+\tfrac{i\hbar a}{k'M\omega_\perp}+\ldots\right)
\end{equation}
 in 1D, as depicted in Fig.~\ref{fig2}. As a consequence, the collision rate will have a a different dependence on the density.  In this paper we show, that different from the 3D case \cite{Ebling2014}, the growth of the collision rate with increasing density is slower than the linear growth of mean-field effects in accordance with the well known lore that in 1D, interaction-induced correlations become stronger for lower densities.

The collision integral is given by
\begin{widetext}
\begin{align}
\label{eq:9}
C_{ij}(r,p)=&-\int dq\int dq'\sum_{m_2\ldots m_8}\left\lbrace \left[T_{im_2m_3m_4}^\dagger(q,q')-T_{m_2im_3m_4}^\dagger(-q,q')\right]
\left[T_{m_5m_6m_7m_8}(q',q)-T_{m_5m_6m_8m_7}(q',-q)\right]\right.\nonumber\\
&\left.\times\frac12\left(\delta(E_q^{m_7m_8}-E_{q'}^{m_3m_5})+\delta(E_q^{im_2}-E_{q'}^{m_5m_6})\right)\right.\nonumber\\
&\left.\times S_{m_3m_5m_4m_6}(r,p-\tfrac12(q-q'),p-\tfrac12(q+q'))W_{m_7j}(r,p)W_{m_8m_2}(r,p-q)\right.\nonumber\\
&\left.-\left[T_{im_2m_3m_5}(q,q')-T_{m_2im_3m_4}(-q,q')\right]
\left[T_{m_5m_6m_7m_8}^\dagger(q',q)-T_{m_5m_6m_8m_7}^\dagger(q',-q)\right]\right.\nonumber\\
&\left.\times\delta(E_{q'}^{m_5m_6}-E_q^{jm_2})S_{jm_7m_2m_8}(r,p,p-q)W_{m_3m_5}(r,p-\tfrac12(q-q'))W_{m_4m_6}(r,p-\tfrac12(q+q'))+\text{h.c.}\right\rbrace,
\end{align}
\end{widetext}
with its derivation detailed in the appendix. The collision integral is derived from the BBGKY hierarchy, which we reduce first to a two-particle level but take into account three-body quantum degeneracy effects, where a third Fermion can occupy and block an outgoing scattering state \cite{Boercker1979}. Then, in order to obtain a closed kinetic equation for the single-particle Wigner function, we approximate the two-particle density matrix as a function of the single particle density-matrix and perform a semi-classical gradient expansion in phase-space \cite{Lhuillier1982I,Lhuillier1982II,Jeon1988,Jeon1989,Mullin1992,Fuchs2003,Ebling2014}. 

In the collision integral, Eq.~\ref{eq:9}, we modify the T-matrix to be explicitly spin-dependent, 
\begin{align}
\label{eq:10}
T_{m_1m_2m_3m_4}(q,q')=&\frac{\hbar}{2\pi M}\frac{i|q'|}{1-\frac{i|q'|2\hbar^2}{M\omega_\perp a_S}}\nonumber\\
&\times\sum_{SM}\langle m_1m_2|SM\rangle\langle SM|m_3m_4\rangle.
\end{align}
The quantum degeneracy effects that arise from taking into account two-body scattering in the presence of a third particle appear in Eq.~\ref{eq:9}
as the shielding factor 
\begin{align}
\label{eq:11}
S_{ijkl}(x,p,q)=\left(\delta_{ij}-W_{ij}(x,p)\right)\left(\delta_{kl}-W_{kl}(x,q)\right)\nonumber\\
-W_{ij}(x,p)W_{kl}(x,q).
\end{align}
 For spin-changing collisions, the Zeeman energy of a pair of atoms changes which is taken into account by the explicit spin dependence of single-particle energies $E_q^{ij}=\frac{q^2}{2M}+Q(i^2+j^2)$.

\begin{figure}[t]
\centering
\includegraphics[width=8.6cm]{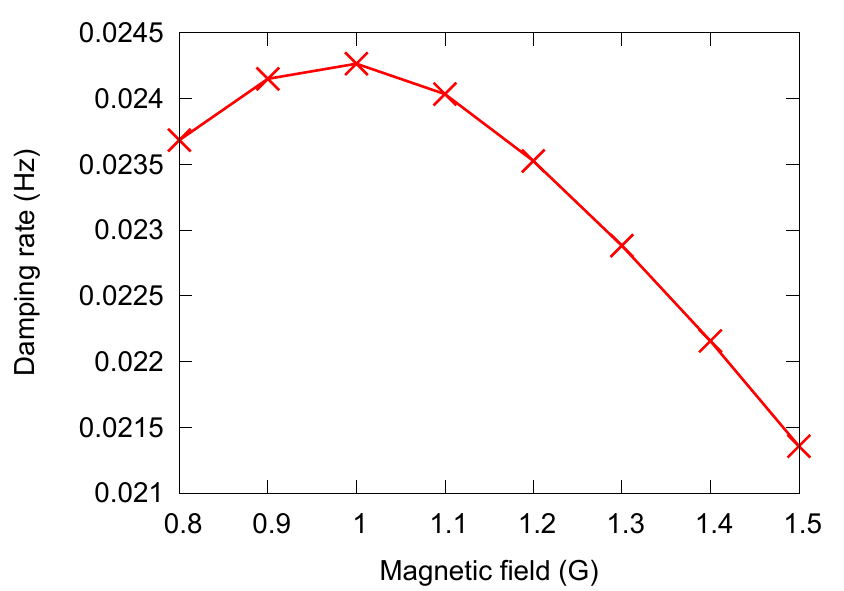}
\caption{Magnetic field dependence of the damping rate of a large-spin 1D Fermi gas. It shows that as expected, damping rates are largely unaffected by the magnetic field, but due to the general non-linear nature of the kinetic equation (\ref{eq:9}), there is a slight dependence of damping rate on frequency and amplitude.}
\label{fig3}
\end{figure}

\begin{figure}[t]
\centering
\includegraphics[width=8.6cm]{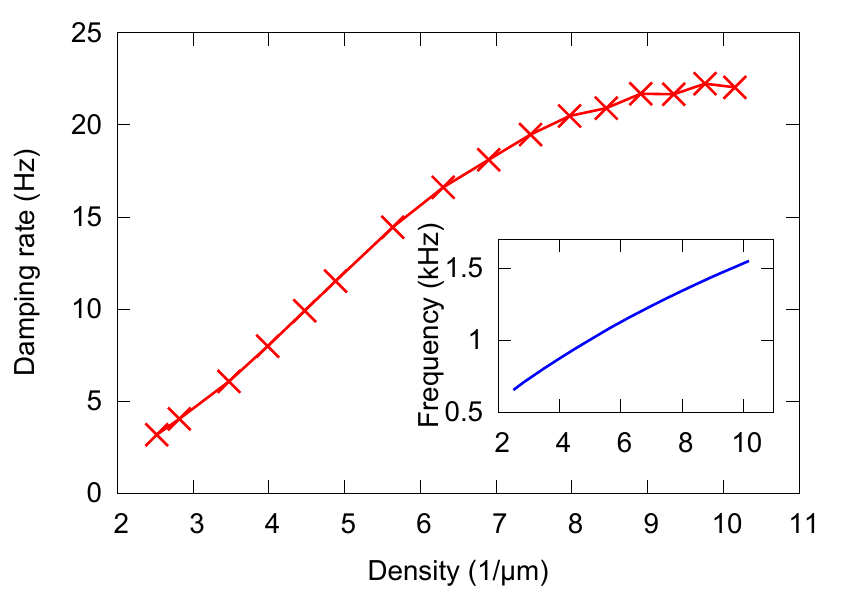}
\caption{Density dependence of the damping rate of coherent spin oscillations between spin populations of the $m=\pm 1/2$ and $m=\pm 3/2$ components in a 1D geometry. Different from the 3D case, the damping rate shows a sub-linear growth with density, related to the dominating large-$k'$ tail of the imaginary part of the T-matrix shown in Fig.~\ref{fig2} and the different density of states for a 1D harmonic trap. The inset shows the bare mean-field oscillation frequency $\omega_\text{mf}$.}
\label{fig4}
\end{figure}

\section{Spin Dynamics}
We can solve equation (\ref{eq:4}) with the full collision integral Eq.~(\ref{eq:9}) numerically for various parameters and initial spin configurations. The initial Wigner function at $t=0$ we consider is given as  
\begin{equation}
\label{eq:12}
W_{ij}(x,p)=\frac{M_{ij}}{2\pi\hbar}\frac1{\exp\left[\frac{1}{k_BT}\left\lbrace\frac{p^2}{2M}+\frac{M\omega^2 x^2}{2}-\mu\right\rbrace\right]+1},
\end{equation}
 where the phase space distribution is approximated by that of a two-component non-interacting Fermi gas, and the matrix with elements $M_{ij}$ is a spin rotation of a two component mixture of hyperfine states states $m=\pm 1/2$ about the x-axis with an angle $\theta$, which corresponds to the application of an rf-pulse. The total particle number is given by $N=\sum_i\int dx\int dp W_{ii}(x,p)$. In this paper, we restrict ourselves to spin-nematic initial states, which are easy to prepare experimentally by e.g.\ trapping a gas of \textsuperscript{40}K atoms in a balanced mixture of hyperfine states $|F=9/2,m=\pm1/2\rangle$ and applying a spin rotation via an rf-pulse \cite{Krauser2014}. The typical time-evolution of such a state is shown in Fig.~\ref{fig1} for a model 4-component ($F=3/2$) system, where we use the scattering lengths $a_0=119.9\,a_B\,a_2=147.8\,a_B$, Zeeman splitting and mass of \textsuperscript{40}K \cite{Krauser2012}. The processes seen, coherent spin oscillations, damping and a long-term redistribution among the spin states have been discussed, as well as their origins, in Ref.~\cite{Ebling2014}, with a focus on the redistribution. Oscillations are caused by spin-changing forward scattering, a mean-field effect, whereas damping and redistribution are collisional effects, captured by spin-conserving and spin-changing terms of the collision integral Eq.~\ref{eq:9}, respectively. Our trapping parameters are $\omega=2\pi\times 100\:\text{Hz}$ and $\omega_\perp=2\pi\times 20\:\text{kHz}$, such that for low temperatures, we can assume transverse modes to be unpopulated.

We first investigate the dependence of coherent spin oscillations and damping on the applied external magnetic field. Since we assume the damping to be driven by spin-conserving collisions, it should not be affected by the magnetic field. However, since the kinetic equation (\ref{eq:4}) is non-linear, oscillation frequency, amplitude and damping may depend on one another. In Fig.~\ref{fig1}, we compare a relatively weak magnetic field with a strong one. As expected, the frequency increases with $B$, while the amplitude decreases, easily understood by viewing the system as a two-level system, where $m=\pm 1/2$ and $m=\pm 3/2$ correspond to ground and excited state, with energy difference $\Delta=4Q$ determined by the QZE and the coupling by the spin-changing scattering length. Another difference is that for high $B$, the oscillations are perfectly harmonic, but not for low $B$. This is due to the appearance of spatial dephasing at low $B$, where the oscillation frequency is dominated by interactions $\sim n(x) g$, as well as the fact that the interaction term in the kinetic equation Eq.~(\ref{eq:4}) is non-linear, while the Zeeman effect is linear. We have obtained the damping rate depicted in Fig. 3, by fitting an exponentially damped harmonic oscillation to the numerical results for the population of components $m=\pm1 /2$, such as depicted in Fig.~\ref{fig1}, for different values of $B$. Below $0.6$ Gauss, the time evolution is no longer described by such a simple expression: Oscillations are no longer harmonic and the damping not exponential, which we attribute the the reasons mentioned above. Therefore, in the remainder of this paper, we set the magnetic field to $B=0.8\,\text{G}$ in order to avoid complex unharmonic dynamics, but keeping the oscillation amplitude and damping rate large.


\section{Coherent versus incoherent processes} 
After this preparatory study, we investigate the dependence of the damping rate of coherent spin oscillations on density, at the same time keeping the temperature constant at $T=0.1\,T_f$, where $T_f$ denotes the Fermi temperature of the two-component Fermi gas prepared before the initial rf-pulse. In Fig.~\ref{fig4}, we depict the damping rate of coherent spin oscillations as a function of the (peak) density, which is varied by changing only the particle number from $40$ to $400$ to ensure the validity of the 1D approximation. We extract it by fitting a damped harmonic oscillation $f(t)=a+b\cos(\omega t+\phi)\exp(-\gamma t)$ into the time evolution of the spin populations as seen in Fig.~\ref{fig4}. We see the expected sub-linear growth of the damping rate for higher densities. Compared to the linear growth of the mean-field oscillation frequency $\hbar\omega_\text{mf}=\sqrt{ \Delta^2-(n(x)g)^2}$ (inset), it means that in 1D for higher densities, mean-field effects such as the spin oscillations become more dominant than dissipative dynamics, opposite to the 1D case.

\begin{figure}[t]
\centering
\includegraphics[width=8.6cm]{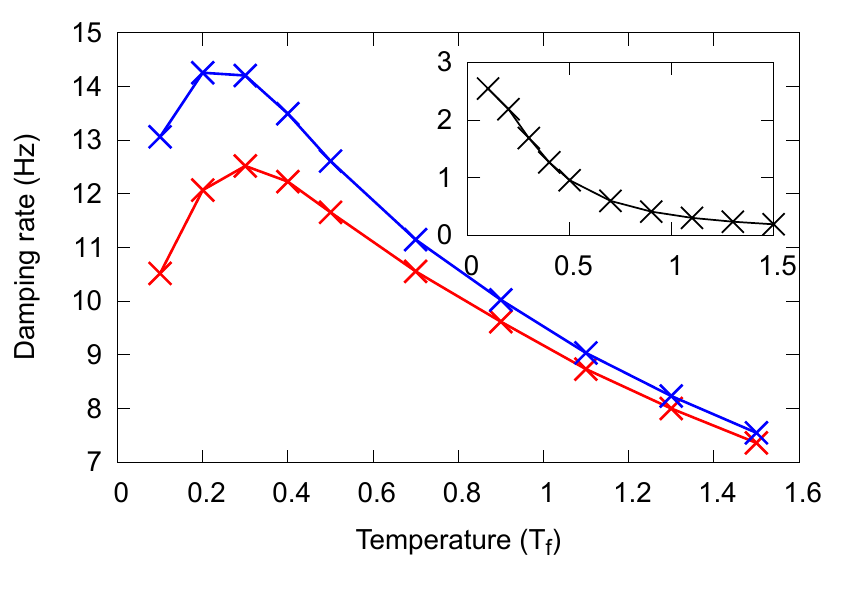}
\caption{Dependence of the damping rate on temperature in units of the Fermi temperature. The red line is obtained using the quantum Boltzmann equation with collision term Eq.~(\ref{eq:9}), while for the blue line, the shielding factors (\ref{eq:11}) that arise from three-body correction are omitted. This shows, that for low temperatures, such degeneracy effects should not be neglected for spin-conserving collisions between Fermi seas with large populations. The inset shows the difference between both cases.}
\label{fig5}
\end{figure}

As a next step, we vary the temperature with respect to the Fermi temperature and keep the density constant, again by changing the particle number, in order to prevent the effect of having increasingly low densities at high temperatures from playing a role, which would make it hard to distinguish density from temperature effects. As depicted in Fig.~\ref{fig5}, the damping rate shows a maximum and decreases both for low and high densities. The drop of the collision rate for high temperatures can be explained by looking again at the momentum dependence of the T-Matrix in Fig.~\ref{fig2}. At high temperatures, the expansion of the gas means, that the large-$k'$ tail of the T-matrix becomes more dominant, although the peak density at $x=0$ does not decrease. This leads to an effective reduction of the collision rate for high temperatures.

At low temperatures, three-body (shielding) effects will become important when when investigating effects driven by spin-conserving collisions. When for low $T$, the number of holes in the Fermi sea decreases, less states are available for particles coming out of collisions, leading to a reduced collision rate. The blue line in Fig.~\ref{fig6} shows the damping rate in case the shielding factor (\ref{eq:11}) is neglected in Eq.~(\ref{eq:9}) and diverges considerably at low temperatures. 
\begin{figure}[t]
\centering
\includegraphics[width=8.6cm]{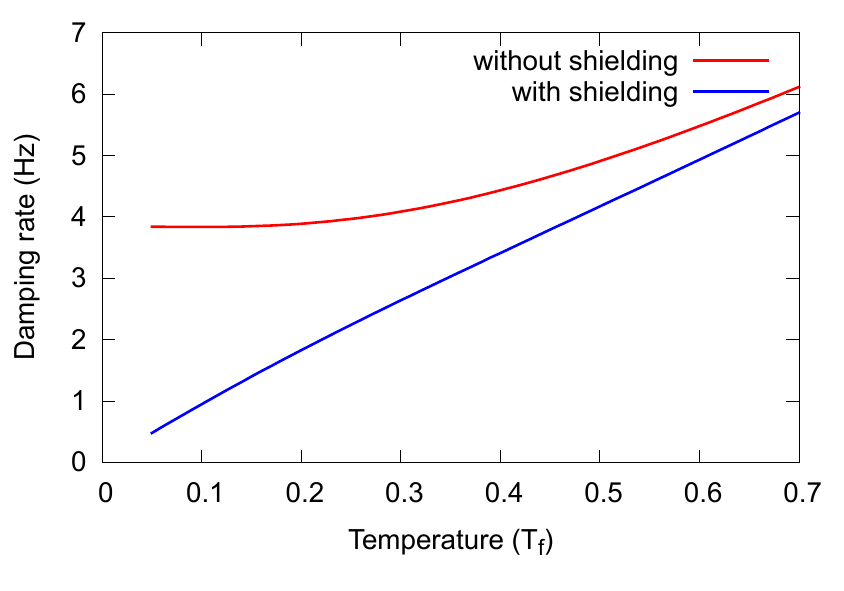}
\caption{Damping rates for spin oscillations in a 3D system of \textsuperscript{40}K, obtained from a 3D Boltzmann equation with a linearized collision term analogous to Eq.~(\ref{eq:9}). Trap frequencies are $\vec\omega=2\pi\times(33,33,137)\,\text{Hz}$, particle numbers in the range of $57000\leq N\leq 473000$ are changed in order to keep the peak density constant, like in the 1D case Fig.~\ref{fig5}. While the difference at low temperatures between the cases with and without the shielding factors is also present and very pronounced, there is no sudden drop in the damping rate because of the different low energy asymptotic behavior of the T-matrix as depicted in Fig.~\ref{fig2}.}
\label{fig6}
\end{figure}

\section{Conclusion}
We investigated the interplay of mean-field and collision driven dynamics of an ultracold spinor Fermi gas in a 1D geometry. We derived a multicomponent quantum Boltzmann equation that takes into account different two-body collision processes as well as degeneracy effects arising from the presence of a third particle not directly involved in the collision, but possibly occupying a final collision state at very low temperatures. With this method, we studied the interplay of a dynamical mean-field effect, coherent spin oscillations, and the damping of these oscillations, caused by spin-conserving collisions which appear as dissipative dynamics on the single-particle level. We looked at these effects for different densities, temperatures and magnetic fields to note behavior for which the behavior is different in a 1D system compared to the usual 3D case. The collision rate features a sub-linear increase with density, different from the 3D case and consistent with the density of states for a 1D harmonic trap. This means, that in 1D gases, dissipative effects such as collision induced damping are expected to become more important compared to mean-field effects when the density is low. Thus, the enhancement of two-particle correlations in 1D for low densities can be studied experimentally by observing the spin dynamics of a large-spin Fermi gas. Our results emphasize that the behavior of dissipative processes, such as relaxation or thermalization, depends on the dimensionality of the system, and conversely that preparing a quantum system with a certain dimensionality can be used to tune such processes.

\begin{acknowledgments}
The authors would like to thank C.~Becker, N.~Fl\"aschner, J.~Heinze, J.S.~Krauser, M.~Lewenstein and K.~Sengstock for useful discussions. This work was supported by the German Research Foundation (DFG) via the Research Unit FOR 2414. U.~E. acknowledges support from a postdoctoral fellowship of the Japan Society for Promotion of Science (JSPS).
\end{acknowledgments}

\appendix
\section{Derivation of the quantum Boltzmann equation}
We derive the quantum Boltzmann equation according to the method described by Jeon and Mullin \cite{Jeon1988,Jeon1989,Mullin1992}, which itself is a generalization of the Snider kinetic equation \cite{Laloe1990} for degenerate gases, obtained by including a ``shielding factor'' to the operator describing two-body collision, such that three-body effects -- the occupation of final states by a third particle in a two-body collision  -- are taken into account. This shielding factor was derived by Boercker and Dufty very generally for degenerate quantum gases \cite{Boercker1979} and later applied to systems of liquid Helium by Jeon and Mullin, and here we shortly reiterate the main ideas of these works. Later, we extend this approach to include multicomponent systems with both spin-conserving and spin-changing collisions.

We consider the a dilute gas of $N$ particles, which we for now treat as distinguishable. The free motion of the $i$-th particle in an external potential is described by Hamiltonian $H(i)$ (e.g. the first row of Eq.~(\ref{eq:1})), and since we consider a dilute system we take into account only two-body interactions described by the operator $V(ij)$, such that the total $N$-body Hamiltonian is given by
\begin{equation}
\label{eq:a1}
H_N=\sum_i H(i)+\frac12\sum_{i\neq j}V(ij).
\end{equation}
The time evolution of that system can be described by the BBGKY hierarchy \cite{Bogoliubov1967lectures}, whose first equations are
\begin{align}
\label{eq:a2}
 i\hbar\frac{d}{dt}\rho(1)=&\left[H(1),\rho(1)\right]+\text{Tr}_2\left\lbrace\left[V(12),\rho(12)\right]\right\rbrace\\
\label{eq:a3}
 i\hbar\frac{d}{dt}\rho(12)=&\left[H(12),\rho(12)\right]\nonumber\\
&+\text{Tr}_3\left\lbrace\left[V(13)+V(23),\rho(123)\right]\right\rbrace
\end{align}
Here, $\rho(1,2,\ldots,N)$ denotes the $N$-body density matrix for \emph{distinguishable} particles $1,2,\ldots,N$ and $\text{Tr}_i$ the trace over all degrees of freedom of the $i$-th particle. To obtain a kinetic equation for just the single-particle density matrix, the hierarchy must be closed by applying appropriate approximations, since the dynamics of the $n$-body density matrix depends on the $n+1$-body density matrix. To take into account the presence of a third particle during the interaction of two other particles, we neglect correlations of the third particle with the other two and approximate the three-body density matrix as
\begin{equation}
\label{eq:a4}
V(12)\rho(123)\approx V(12)\rho(12)\rho(3),
\end{equation}
with the argument that for short-range interactions, the third particle is unlikely to be close enough to the other two to be inside the range of the interaction potential. This procedure closes the BBGKY-hierarchy for the two-body density matrix. To switch from distinguishable to indistinguishable particles and obtain proper (anti-)symmetrization of the wavefunctions, an exchange operator $\mathcal P_{ij}$ is introduced, which exchanges all degrees of freedom of particles $i$ and $j$, e.g.
\begin{equation}
\label{eq:a5}
\mathcal P_{12} \rho(123)\mathcal P_{12}=\rho(213).
\end{equation} 
With the approximation (\ref{eq:a5}) and the inclusion of particle indistinguishability, Eqs.~(\ref{eq:a2},\ref{eq:a3}) become 
\begin{align}
\label{eq:a6}
 i\hbar\frac{d}{dt}\rho(1)&=\left[H(1),\rho(1)\right]+\text{Tr}_2\left\lbrace\left[V(12),\rho(12)\right]\right\rbrace\\
\label{eq:a7}
 i\hbar\frac{d}{dt}\rho(12)=&\left[H(12),\rho(12)\right]\nonumber\\
&+\left[S(12)V(12)\rho(12)-\rho(12)V(12)S(12)\right]\nonumber\\
&+\frac{i\hbar}{2}\frac{d}{dt}(1+\epsilon\mathcal P)\rho(1)\rho(2)(1+\epsilon\mathcal P)\nonumber\\
&-\frac12\left[H(12),(1+\epsilon\mathcal P)\rho(1)\rho(2)(1+\epsilon\mathcal P)\right],
\end{align}
where $\mathcal P\equiv \mathcal P_{12}$, $H(12)=H(1)+H(2)$. Most importantly, the effect of the third particle appears in the shielding factor 
\begin{equation}
\label{eq:a8}
S(12)=\tilde\rho(1)\tilde\rho(2)-\rho(1)\rho(2),
\end{equation}
where
\begin{equation}
\label{eq:a9}
 \tilde\rho(i)=1+\epsilon\rho(i)
\end{equation}
depends on whether our system is bosonic ($\epsilon=1$) or fermionic ($\epsilon=-1$). Note that for fermions $\tilde{\rho}$ corresponds to the density of holes.

The further reduction of (\ref{eq:a6}) and (\ref{eq:a7}) to a kinetic equation for just the single-particle density matrix is shown rigorously and in full detail in \cite{Boercker1979}, it involves the formal solution of Eq.~(\ref{eq:a7}) with $\rho(12)$ treated as a functional of the single-particle density matrix. It is then shown that for dilute gases with short-range interactions, this procedure is equivalent to making the Ansatz
\begin{equation}
\label{eq:a10}
 \rho(1,2)\approx\frac12(1+\epsilon\mathcal P)\Omega(12)\rho(1)\rho(2)\Omega(12)^\dagger(1+\epsilon\mathcal P),
\end{equation}
for the two-body density matrix in the first equation of the BBGKY hierarchy. Here, $\Omega(12)$ denotes the M\o ller wave operator associated with the interaction potential $V(1,2)$. This operator effectively connects the two body density matrices before and after a collision: $\rho(12)=\Omega(12)\rho(12)|_\text{initial}\Omega(12)^\dagger$ and the approximation made here, dating back to work of Snider \cite{Snider1960,Thomas1970}, is to assume particles before the collision to be uncorrelated: $\rho(12)|_\text{initial}\approx\rho(1)\rho(2)$.

With all these considerations, the first equation of the BBGKY hierarchy becomes a closed equation of motion for the single particle density matrix
\begin{align}
\label{eq:a11}
  i\hbar\frac{d}{dt}\rho(1)&=\left[H(1),\rho(1)\right]\nonumber\\
	&+\frac12\text{Tr}_2\left\lbrace(1+\epsilon\mathcal P)\left[V,\Omega\rho(1)\rho(2)\Omega^\dagger\right](1+\epsilon\mathcal P)\right\rbrace,
\end{align}
where $V\equiv V(12)$ and $\Omega\equiv \Omega(12)$.

The M\o ller operator is related to the T-matrix via the identity
\begin{equation}
\label{eq:a12}
 T=V\Omega,
\end{equation}
such that the commutator in equation (\ref{eq:a11}) now becomes
\begin{align}
\label{eq:a13}
 \left[V,\Omega\rho(1)\rho(2)\Omega^\dagger\right]&=V\Omega\rho(1)\rho(2)\Omega^\dagger-\Omega\rho(1)\rho(2)\Omega^\dagger V\nonumber\\
 &=T\rho(1)\rho(2)\Omega^\dagger-\Omega\rho(1)\rho(2)T^\dagger.
\end{align}
A comparison of the Ansatz (\ref{eq:a10}) with the method of closing the BBGKY hierarchy developed by Boercker and Dufty \cite{Boercker1979}, yield that the T-matrix for two-body scattering  fulfills the identity
\begin{equation}
\label{eq:a14}
 T=V+VR(E)T,
\end{equation}
where the resolvent is given as
\begin{equation}
\label{eq:a15}
 R(E)=\frac{1}{E+i0^{+}-H(12)}S(12)
\end{equation}
and takes into account the presence of a third particle with the shielding factor (\ref{eq:a8}). This factor reduces or enhances the scattering amplitude, depending on particle statistics, e.g. in the fermionic case a two-body collision is forbidden if one of the outgoing states is already occupied by a third body.

Combining equations (\ref{eq:a12}) and (\ref{eq:a14}), we obtain
\begin{equation}
\label{eq:a16}
 \Omega=1+R(E)T,
\end{equation}
which we substitute into equation (\ref{eq:a11}) to arrive at the expression
\begin{align}
\label{eq:a17}
  \left[V,\Omega\rho(1)\rho(2)\Omega^\dagger\right]=T\rho(1)\rho(2)-\rho(1)\rho(2)T^\dagger\nonumber\\
	+T\rho(1)\rho(2)T^\dagger R^\dagger-RT\rho(1)\rho(2)T^\dagger.
\end{align}
The resulting kinetic equation now takes the form of a Boltzmann equation
\begin{align}
\label{eq:a18}
  \frac{d}{dt}\rho(1)=&\frac{1}{i\hbar}\left[H(1),\rho(1)\right]\nonumber\\
	&+\frac1{2i\hbar}\text{Tr}_2\left\lbrace(1+\epsilon\mathcal P)\left(T\rho(1)\rho(2)-\rho(1)\rho(2)T^\dagger\right.\right.\nonumber\\
	&\left.\left.+T\rho(1)\rho(2)T^\dagger R^\dagger-RT\rho(1)\rho(2)T^\dagger\right)(1+\epsilon\mathcal P)\right\rbrace
\end{align}
and contains two terms linear and quadratic in the T-matrix each. It has been shown, that for weak interactions, the real part of the T-matrix is approximately linear in the scattering length, dominates and describes forward-scattering so it can be treated on a mean-field level, while the imaginary part is quadratic in scattering lengths and corresponds to lateral collisions \cite{Lhuillier1982I,Lhuillier1982II,Mullin1992,Fuchs2003,Ebling2014}. Hence, at this stage we split up the linear terms into real and imaginary contributions
\begin{align}
\label{eq:a19}
 T&=\tfrac12(T+T^\dagger)+\tfrac12(T-T^\dagger)\nonumber\\
 T^\dagger&=\tfrac12(T+T^\dagger)-\tfrac12(T-T^\dagger)
\end{align}
and make use of the optical theorem \cite{Boercker1979} to transform the imaginary parts into quadratic terms
\begin{equation}
\label{eq:a20}
T-T^\dagger=T^\dagger(R-R^\dagger)T. 
\end{equation}
We now separate the mean-field terms in equation (\ref{eq:a18}) and transform it into
\begin{equation}
\label{eq:a21}
 \frac{d}{dt}\rho(1)-\frac{1}{i\hbar}\left[H(1),\rho(1)\right]-\frac{1}{i\hbar}\left[V_\text{mf},\rho(1)\right]=I_\text{coll}[\rho(1)].
\end{equation}
It consists of three contributions, from left to right: the drift term of the atoms moving freely in the harmonic trap and external magnetic field, a mean-field interaction term which we derive from the real part of the T-matrix
\begin{align}
\label{eq:a22}
 \left[V_\text{mf},\rho(1)\right]=&\frac14\text{Tr}_2\left\lbrace(1+\epsilon\mathcal P)\left((T+T^\dagger)\rho(1)\rho(2)\right.\right.\nonumber\\
&\left.\left.-\rho(1)\rho(2)(T+T^\dagger)\right)(1+\epsilon\mathcal P)\right\rbrace
\end{align}
and the collision integral
\begin{align}
\label{eq:a23}
 I_\text{coll}[\rho(1)]=&\frac1{2i\hbar}\text{Tr}_2\left\lbrace(1+\epsilon\mathcal P)\left(\tfrac12T^\dagger(R-R^\dagger)T\rho(1)\rho(2)\right.\right.\nonumber\\
&\left.\left.+\tfrac12\rho(1)\rho(2)T^\dagger(R-R^\dagger)T+T\rho(1)\rho(2)T^\dagger R^\dagger\right.\right.\nonumber\\
&\left.\left.-RT\rho(1)\rho(2)T^\dagger\right)(1+\epsilon\mathcal P)\right\rbrace.
\end{align}
The mean-field term does not contain the shielding factors for the presence of a third body at the collision, since they describe forward scattering, where the individual momenta of incoming and outgoing particles are unchanged, and consequently the final states cannot be occupied by a third particle. This term has been derived in earlier work \cite{Heinze2013,Krauser2014,Ebling2014} and is given by the mean-field potential
\begin{equation}
\label{eq:a24}
 V^\text{mf}_{ij}(x)=2\sum_{kl}\int dp U_{ijlk} W_{kl}(x,p)
\end{equation}
and coupling constants
\begin{equation}
\label{eq:a25}
 U_{klmn}=\sum_{SM}g_S\langle km|SM\rangle\langle SM|ln\rangle.
\end{equation}

The Wigner representation is the phase space representation of the single-particle density matrix
\begin{equation}
\label{eq:a26}
\rho_{mn}(p,q)\equiv\langle p,m |\rho(1)| q,n\rangle,
\end{equation}
related by the Wigner transform
\begin{equation}
W_{mn}(x,p)=\int dq e^{\tfrac{iqx}{\hbar}}\rho_{mn}(p+\tfrac q2,p-\tfrac q2)
\end{equation}
anf its inverse
\begin{equation}
\rho_{mn}(p,q)=\frac{1}{2\pi\hbar}\int dx e^{\tfrac{ix(p-q)}{\hbar}}W_{mn}(x,\tfrac{p+q}{2}).
\end{equation}
Since in the expression above for the collision integral, Eq.~(\ref{eq:a23}), the second and fourth terms are the hermitian conjugate of their preceding term, we explicitly derive only the contributions arising from these terms. We start with $C^1=\tfrac1{2i\hbar}\text{Tr}_2(T^\dagger(R-R^\dagger)T\rho(1)\rho(2)$, which in its Wigner representation reads
\begin{align}
\label{eq:a27}
 C^1_{ij}(r_1,p_1)&=\frac{1}{4 i\pi\hbar^2}\int dq_1\int dq_2\int d p_2\int dr_2\sum_{m_2}\nonumber\\
&\times e^{\tfrac i\hbar q_1 r_1}e^{\tfrac i\hbar q_2 r_2}\langle p_1^{+} i, p_2^{+}m_2|T^\dagger(R-R^\dagger)\nonumber\\
&\times T\rho(1)\rho(2)| p_1^{-}j, p_2^{-}m_2\rangle,
\end{align}
where $p_i^{\pm}= p_i\pm q_i/2$. We now insert three complete bases $\int dp_i\int dp_j\sum_{m_i,m_j}\left|p_i m_i,p_j m_j\rangle\langle p_i m_i,p_j m_j\right|$ such that
\begin{align}
\label{eq:a28}
 C^1_{ij}(r_1,p_1)&=\frac{1}{4 i\pi\hbar^2}\int dq_1\int dq_2\idotsint d p_{2\ldots 8}\int d r_2\nonumber\\
&\times e^{\tfrac i\hbar q_1 r_1}e^{\tfrac i\hbar q_2 r_2}\nonumber\\
&\times\sum_{m_2\ldots m_8}\langle p_1^{+} i, p_2^{+}m_2|T^\dagger|p_3 m_3,p_4 m_4\rangle\nonumber\\
 &\times\langle p_3 m_3,p_4 m_4|R-R^\dagger|p_5 m_5,p_6 m_6\rangle\nonumber\\
&\times\langle p_5 m_5, p_6 m_6|T|p_7 m_7,p_8 m_8\rangle\nonumber\\
&\times\langle p_7 m_7,p_8 m_8|\rho(1)\rho(2)|p_1^{-}j,p_2^{-}m_2\rangle.
\end{align}
Because the T-matrix depends only on relative momenta, we substitute
\begin{align}
\label{eq:a29}
\langle p_1 m_1, p_2 m_2|T|p_3 m_3,p_4 m_4\rangle&=\delta(p_1+p_2-p_3-p_4)\nonumber\\
&\times T_{m_1 m_2 m_3 m_4}(p_{12},p_{34}), 
\end{align}
where $p_{ij}=\tfrac12(p_i-p_j)$. Energy conservation during a collision is taken care of by the resolvent operator, and we keep the on-shell terms derived in \cite{Boercker1979}, which means that the product $T^\dagger R^\dagger$ in Eq.~(\ref{eq:a28}) is given by
\begin{align}
\label{eq:a30}
 \langle p_1 m_1,&p_2 m_2|T^\dagger|p_3 m_3,p_4 m_4\rangle\langle p_3 m_3,p_4 m_4|R^\dagger|p_5 m_5,p_6 m_6\rangle\nonumber\\
&=i\pi\delta(E_{12}^{m_1 m_2}-E_{56}^{m_5 m_6})\delta(p_1+p_2-p_3-p_4)\nonumber\\
 &\times\left[\delta_{m_3 m_5}\delta_{m_4 m_6}-\delta_{m_3 m_5}\rho_{m_4 m_6}(p_4,p_6)\right.\nonumber\\
&\left.-\delta_{m_4 m_6}\rho_{m_3 m_5}(p_3,p_5)\right]T_{m_1m_2m_3m_4}^\dagger(p_{12},p_{34}),
\end{align}
with two-particle energies
\begin{equation}
\label{eq:a31}
 E_{ij}^{m_1 m_2}=\frac{p_i^2+p_j^2}{2\mu}+Q(m_1^2+m_2^2).
\end{equation}
With these substitutions, Eq.~(\ref{eq:a28}) becomes
\begin{align}
\label{eq:a32}
 C^1_{ij}(r_1,p_1)&=\frac{1}{4\hbar^2}\int dq_1\int dq_2\idotsint d p_{2\ldots 8}\int d r_2\nonumber\\ 
&\times e^{\tfrac i\hbar q_1 r_1}e^{\tfrac i\hbar q_2 r_2}\delta(p_1^{+}+p_2^{+}-p_3-p_4)\sum_{m_2\ldots m_8}\nonumber\\
&\times\delta(p_5+p_6-p_7-p_8)\delta(E_{1+2+}^{12}-E_{56}^{m_5 m6})\nonumber\\
&\times T^\dagger_{im_2m_3m_4}(p_{12}^{+},p_{34})T_{m_5m_6m_7m_8}(p_{56},p_{78})\nonumber\\
&\times\left[\delta_{m_3 m_5}\delta_{m_4 m_6}-\delta_{m_3 m_5}\rho_{m_4 m_6}(p_4,p_6)\right.\nonumber\\
&\left.-\delta_{m_4 m_6}\rho_{m_3 m_5}(p_3,p_5)\right]\nonumber\\
&\times\rho_{m_7j}(p_7,p_1^{-})\rho_{m_8m_2}(p_8,p_2^{-}).
\end{align}

In semi-classical approximation, this expression simplifies considerably. We assume the Wigner function (\ref{eq:3}) to vary only slowly in space compared to the relevant single-particle wavelengths in the system, such that we approximate non-local contributions as
\begin{equation}
\label{eq:a33}
 W_{mn}(r_1+y,p)=W_{mn}(r_1,p)+\ldots,
\end{equation}
which leads to the occurrence of more delta-functions in expression (\ref{eq:a30}), because 
\begin{align}
\label{eq:a34}
 \rho_{mn}(p,q)&=\frac{1}{2\pi\hbar}\int dy e^{\tfrac{i(r_1+y)(p-q)}{\hbar}}W_{mn}(r_1+y,\tfrac{p+q}{2})\nonumber\\
 &=\delta(p-q)W_{mn}(r_1,\tfrac{p+q}{2}).
\end{align}
After carrying out the respective integrations, we obtain
\begin{align}
\label{eq:a35}
 C^1_{ij}(r_1,p_1)&=-\frac{1}{2i\hbar}\int dp_2\int dp_3\int dp_4\sum_{m_2\ldots m_8}\nonumber\\
&\times T_{im_2m_3m_4}^\dagger(p_{12},p_{34})T_{m_3m_4m_5m_6}(p_{34},p_{12})\nonumber\\
&\times\delta(p_1+p_2-p_3-p_4)S_{m_3m_5m_4m_6}(r_1,p_3,p_4)\nonumber\\
 &\times\left[i\pi\delta(E_{12}^{m_5m_6}-E_{34}^{m_3m_4})\right.\nonumber\\
&\left.+i\pi\delta(E_{12}^{im_2}-E_{34}^{m_5m_6})\right]\nonumber\\
&\times W_{m_7j}(r_1,p_1)W_{m_8m_2}(r_1,p_2).
\end{align}
with the Wigner representation of the shielding factor that contains the effect of a third particle present during a binary collision:
\begin{align}
\label{eq:a36}
 S_{m_3m_5m_4m_6}(r_1,p_3,p_4)&=\tilde W_{m_3m_5}(r_1,p_3)\tilde W_{m_4m_6}(r_1,p_4)\nonumber\\
&-W_{m_3m_5}(r_1,p_3)W_{m_4m_6}(r_1,p_4)\nonumber\\
&=\delta_{m_3m_5}\delta_{m_4m_6}\nonumber\\
&-\delta_{m_3m_5}W_{m_4m_6}(r_1,p_4)\nonumber\\
&-\delta_{m_4m_6}W_{m_3m_5}(r_1,p_3).
\end{align}
With an analogous calculation for the third term in Eq.~(\ref{eq:a23}),
\begin{align}
\label{eq:a37}
 C^3_{ij}(r_1,p_1)&=\frac{1}{2 i\pi\hbar^2}\int dq_1\int dq_2\int dp_2\int dr_2 e^{\tfrac{i}{\hbar}(q_1 r_1+q_2 r_2)}\nonumber\\
 &\times\sum_{m_2}\langle p_1^{+} i,p_2^{+} m_2|T\rho(1)\rho(2)T^\dagger R^\dagger|p_1^{-} j,p_2^{-} m_2 \rangle,
\end{align}
we find 
\begin{align}
\label{eq:a38}
C^3_{ij}(r_1,p_1)&=\frac{1}{i\hbar}\int dp_2\int dp_3\int dp_4\sum_{m_2\ldots m_8}\nonumber\\
&\times T_{im_2m_3m_4}(p_{12},p_{34})T_{m_5m_6m_7m_8}^\dagger(p_{34},p_{12})\nonumber\\
&\times\delta(p_1+p_2-p_3-p_4)\nonumber\\
&\times i\pi\delta(E_{34}^{m_5m_6}-E_{12}^{jm_2})S_{jm_6m_2m_8}(r_1,p_1,p_2)\nonumber\\
&\times W_{m_3m_5}(r_1,p_3)W_{m_4m_6}(r_1,p_4).
\end{align}
We then obtain the full collision integral by adding the hermitian conjugate terms in (\ref{eq:a23}), as well as taking into account the exchange terms we so far neglected, to the results above. The final result is
\widetext
\begin{align}
\label{eq:a39}
 C_{ij}(r,p)=&-\frac12\int dp_2\int dp_3\int dp_4\sum_{m_2\ldots m_8}\delta(p+p_2-p_3-p_4)\left\lbrace
 \left[T_{im_2m_3m_4}^\dagger(p_{12},p_{34})-T_{m_2im_3m_4}^\dagger(-p_{12},p_{34})\right]\right.\nonumber\\
 &\left.
 \times\left[T_{m_5m_6m_7m_8}(p_{34},p_{12})-T_{m_5m_6m_8m_7}(p_{34},-p_{12})\right]
 \frac12\left[\delta(E_{12}^{m_7m_8}-E_{34}^{m_3m_4})+\delta(E_{12}^{im_2}-E_{34}^{m_5m_6})\right]\right.\nonumber\\
 &\left.\times S_{m_3m_5m_4m_6}(r,p_3,p_4) W_{m_7j}(r,p)W_{m_8m_2}(r,p_2)-\left[T_{im_2m_3m_4}(p_{12},p_{34})-T_{m_2im_3m_4}(-p_{12},p_{34})\right]\right.\nonumber\\
&\left.\times\left[T_{m_5m_6m_7m_8}^\dagger(p_{34},p_{12})-T_{m_5m_6m_8m_7}^\dagger(p_{34},-p_{12})\right]
 \delta(E_{34}^{m_5m_6}-E_{12}^{jm_2})\right.\nonumber\\
 &\left.\times S_{jm_7m_2m_8}(r,p,p_2) W_{m_3m_5}(r,p_3)W_{m_4m_6}(r,p_4) +\text{h.c.}
 \right\rbrace.
\end{align}
The delta functions of energy are of the form
\begin{align}
\label{eq:a40}
 \delta(E-E')&=\delta(\tfrac{p^2-q^2+\Delta}{2M})\nonumber\\
 &=\frac{M}{\sqrt{q^2+\Delta}}\left[\delta(p+\sqrt{q^2+\Delta})+\delta(p-\sqrt{q^2+\Delta})\right]\nonumber\\
\end{align}
such that momentum conservation in the presence of a quadratic Zeeman effect appears as e.\ g.\
\begin{align}
\label{eq:a41}
 \delta(E_{34}^{m_3m_4}-E_{12}^{m_1m_2})=&\frac{M}{\sqrt{p_{12}^2+Q(m_1^2+m_2^2-m_3^2-m_4^2)}}\left[\delta\left(p_{34}+\sqrt{p_{12}^2+Q(m_1^2+m_2^2-m_3^2-m_4^2)}\right)\right.\nonumber\\
&\left.+\delta\left(p_{34}-\sqrt{p_{12}^2+Q(m_1^2+m_2^2-m_3^2-m_4^2)}\right)\right].
\end{align}
This allows us to eliminate further integrals, and together with a change of coordinates into relative momenta, we obtain
\begin{align}
\label{eq:a42}
C_{ij}(r,p)=&-\int dq\int dq'\sum_{m_2\ldots m_8}\left\lbrace \left[T_{im_2m_3m_4}^\dagger(q,q')-T_{m_2im_3m_4}^\dagger(-q,q')\right]
\left[T_{m_5m_6m_7m_8}(q',q)-T_{m_5m_6m_7m_8}(q',-q)\right]\right.\nonumber\\
&\left.\times\frac12\left(\delta(E_q^{m_7m_8}-E_{q'}^{m_3m_4})+\delta(E_q^{im_2}-E_{q'}^{m_5m_6})\right)S_{m_3m_5m_4m_6}(r,p-\tfrac12(q-q'),p-\tfrac12(q+q'))\right.\nonumber\\
&\left.\times W_{m_7j}(r,p)W_{m_8m_2}(r,p-q)\right.\nonumber\\
&\left.-\left[T_{im_2m_3m_4}(q,q')-T_{m_2im_3m_4}(-q,q')\right]\left[T_{m_5m_6m_7m_8}^\dagger(q',q)-T_{m_5m_6m_8m_7}^\dagger(q',-q)\right]\right.\nonumber\\
&\left.\delta(E_{q'}^{m_5m_6}-E_q^{jm_2})S_{jm_7m_2m_8}(r,p,p-q)W_{m_3m_5}(r,p-\tfrac12(q-q'))W_{m_4m_6}(r,p-\tfrac12(q+q'))+\text{h.c.}\right\rbrace
\end{align}
for the collision integral. Note that this expression is very similar to the collision integral in \cite{Ebling2014}, however it additionally contains corrections for the degenerate regime in the form of the shielding factor $S$. Indeed, the substitution of $S_{ijkl}(x,p,q)\rightarrow\delta_{ij}\delta_{kl}$ into Eq.~(\ref{eq:a42}), provides the collision term derived in Ref.~\cite{Ebling2014}.

\bibliography{refs}

\begin{thebibliography}{41}%
\makeatletter
\providecommand \@ifxundefined [1]{%
 \@ifx{#1\undefined}
}%
\providecommand \@ifnum [1]{%
 \ifnum #1\expandafter \@firstoftwo
 \else \expandafter \@secondoftwo
 \fi
}%
\providecommand \@ifx [1]{%
 \ifx #1\expandafter \@firstoftwo
 \else \expandafter \@secondoftwo
 \fi
}%
\providecommand \natexlab [1]{#1}%
\providecommand \enquote  [1]{``#1''}%
\providecommand \bibnamefont  [1]{#1}%
\providecommand \bibfnamefont [1]{#1}%
\providecommand \citenamefont [1]{#1}%
\providecommand \href@noop [0]{\@secondoftwo}%
\providecommand \href [0]{\begingroup \@sanitize@url \@href}%
\providecommand \@href[1]{\@@startlink{#1}\@@href}%
\providecommand \@@href[1]{\endgroup#1\@@endlink}%
\providecommand \@sanitize@url [0]{\catcode `\\12\catcode `\$12\catcode
  `\&12\catcode `\#12\catcode `\^12\catcode `\_12\catcode `\%12\relax}%
\providecommand \@@startlink[1]{}%
\providecommand \@@endlink[0]{}%
\providecommand \url  [0]{\begingroup\@sanitize@url \@url }%
\providecommand \@url [1]{\endgroup\@href {#1}{\urlprefix }}%
\providecommand \urlprefix  [0]{URL }%
\providecommand \Eprint [0]{\href }%
\providecommand \doibase [0]{http://dx.doi.org/}%
\providecommand \selectlanguage [0]{\@gobble}%
\providecommand \bibinfo  [0]{\@secondoftwo}%
\providecommand \bibfield  [0]{\@secondoftwo}%
\providecommand \translation [1]{[#1]}%
\providecommand \BibitemOpen [0]{}%
\providecommand \bibitemStop [0]{}%
\providecommand \bibitemNoStop [0]{.\EOS\space}%
\providecommand \EOS [0]{\spacefactor3000\relax}%
\providecommand \BibitemShut  [1]{\csname bibitem#1\endcsname}%
\let\auto@bib@innerbib\@empty
\bibitem [{\citenamefont {G\"orlitz}\ \emph {et~al.}(2001)\citenamefont
  {G\"orlitz}, \citenamefont {Vogels}, \citenamefont {Leanhardt}, \citenamefont
  {Raman}, \citenamefont {Gustavson}, \citenamefont {Abo-Shaeer}, \citenamefont
  {Chikkatur}, \citenamefont {Gupta}, \citenamefont {Inouye}, \citenamefont
  {Rosenband},\ and\ \citenamefont {Ketterle}}]{Goerlitz2001}%
  \BibitemOpen
  \bibfield  {author} {\bibinfo {author} {\bibfnamefont {A.}~\bibnamefont
  {G\"orlitz}}, \bibinfo {author} {\bibfnamefont {J.~M.}\ \bibnamefont
  {Vogels}}, \bibinfo {author} {\bibfnamefont {A.~E.}\ \bibnamefont
  {Leanhardt}}, \bibinfo {author} {\bibfnamefont {C.}~\bibnamefont {Raman}},
  \bibinfo {author} {\bibfnamefont {T.~L.}\ \bibnamefont {Gustavson}}, \bibinfo
  {author} {\bibfnamefont {J.~R.}\ \bibnamefont {Abo-Shaeer}}, \bibinfo
  {author} {\bibfnamefont {A.~P.}\ \bibnamefont {Chikkatur}}, \bibinfo {author}
  {\bibfnamefont {S.}~\bibnamefont {Gupta}}, \bibinfo {author} {\bibfnamefont
  {S.}~\bibnamefont {Inouye}}, \bibinfo {author} {\bibfnamefont
  {T.}~\bibnamefont {Rosenband}}, \ and\ \bibinfo {author} {\bibfnamefont
  {W.}~\bibnamefont {Ketterle}},\ }\href {\doibase
  10.1103/PhysRevLett.87.130402} {\bibfield  {journal} {\bibinfo  {journal}
  {Phys. Rev. Lett.}\ }\textbf {\bibinfo {volume} {87}},\ \bibinfo {pages}
  {130402} (\bibinfo {year} {2001})}\BibitemShut {NoStop}%
\bibitem [{\citenamefont {Greiner}\ \emph {et~al.}(2001)\citenamefont
  {Greiner}, \citenamefont {Bloch}, \citenamefont {Mandel}, \citenamefont
  {H\"ansch},\ and\ \citenamefont {Esslinger}}]{Greiner2001}%
  \BibitemOpen
  \bibfield  {author} {\bibinfo {author} {\bibfnamefont {M.}~\bibnamefont
  {Greiner}}, \bibinfo {author} {\bibfnamefont {I.}~\bibnamefont {Bloch}},
  \bibinfo {author} {\bibfnamefont {O.}~\bibnamefont {Mandel}}, \bibinfo
  {author} {\bibfnamefont {T.~W.}\ \bibnamefont {H\"ansch}}, \ and\ \bibinfo
  {author} {\bibfnamefont {T.}~\bibnamefont {Esslinger}},\ }\href {\doibase
  10.1103/PhysRevLett.87.160405} {\bibfield  {journal} {\bibinfo  {journal}
  {Phys. Rev. Lett.}\ }\textbf {\bibinfo {volume} {87}},\ \bibinfo {pages}
  {160405} (\bibinfo {year} {2001})}\BibitemShut {NoStop}%
\bibitem [{\citenamefont {Laburthe~Tolra}\ \emph {et~al.}(2004)\citenamefont
  {Laburthe~Tolra}, \citenamefont {O'Hara}, \citenamefont {Huckans},
  \citenamefont {Phillips}, \citenamefont {Rolston},\ and\ \citenamefont
  {Porto}}]{Tolra2004}%
  \BibitemOpen
  \bibfield  {author} {\bibinfo {author} {\bibfnamefont {B.}~\bibnamefont
  {Laburthe~Tolra}}, \bibinfo {author} {\bibfnamefont {K.~M.}\ \bibnamefont
  {O'Hara}}, \bibinfo {author} {\bibfnamefont {J.~H.}\ \bibnamefont {Huckans}},
  \bibinfo {author} {\bibfnamefont {W.~D.}\ \bibnamefont {Phillips}}, \bibinfo
  {author} {\bibfnamefont {S.~L.}\ \bibnamefont {Rolston}}, \ and\ \bibinfo
  {author} {\bibfnamefont {J.~V.}\ \bibnamefont {Porto}},\ }\href {\doibase
  10.1103/PhysRevLett.92.190401} {\bibfield  {journal} {\bibinfo  {journal}
  {Phys. Rev. Lett.}\ }\textbf {\bibinfo {volume} {92}},\ \bibinfo {pages}
  {190401} (\bibinfo {year} {2004})}\BibitemShut {NoStop}%
\bibitem [{\citenamefont {Pagano}\ \emph {et~al.}(2014)\citenamefont {Pagano},
  \citenamefont {Mancini}, \citenamefont {Cappellini}, \citenamefont
  {Lombardi}, \citenamefont {Sch\"afer}, \citenamefont {Hu}, \citenamefont
  {Liu}, \citenamefont {Catani}, \citenamefont {Sias}, \citenamefont
  {Inguscio},\ and\ \citenamefont {Fallani}}]{Fallani2014}%
  \BibitemOpen
  \bibfield  {author} {\bibinfo {author} {\bibfnamefont {G.}~\bibnamefont
  {Pagano}}, \bibinfo {author} {\bibfnamefont {M.}~\bibnamefont {Mancini}},
  \bibinfo {author} {\bibfnamefont {G.}~\bibnamefont {Cappellini}}, \bibinfo
  {author} {\bibfnamefont {P.}~\bibnamefont {Lombardi}}, \bibinfo {author}
  {\bibfnamefont {F.}~\bibnamefont {Sch\"afer}}, \bibinfo {author}
  {\bibfnamefont {H.}~\bibnamefont {Hu}}, \bibinfo {author} {\bibfnamefont
  {X.-J.}\ \bibnamefont {Liu}}, \bibinfo {author} {\bibfnamefont
  {J.}~\bibnamefont {Catani}}, \bibinfo {author} {\bibfnamefont
  {C.}~\bibnamefont {Sias}}, \bibinfo {author} {\bibfnamefont {M.}~\bibnamefont
  {Inguscio}}, \ and\ \bibinfo {author} {\bibfnamefont {L.}~\bibnamefont
  {Fallani}},\ }\href {\doibase 10.1038/nphys2878} {\bibfield  {journal}
  {\bibinfo  {journal} {Nature Physics}\ }\textbf {\bibinfo {volume} {10}},\
  \bibinfo {pages} {198} (\bibinfo {year} {2014})}\BibitemShut {NoStop}%
\bibitem [{\citenamefont {Olshanii}(1998)}]{Olshanii1998}%
  \BibitemOpen
  \bibfield  {author} {\bibinfo {author} {\bibfnamefont {M.}~\bibnamefont
  {Olshanii}},\ }\href {\doibase 10.1103/PhysRevLett.81.938} {\bibfield
  {journal} {\bibinfo  {journal} {Phys. Rev. Lett.}\ }\textbf {\bibinfo
  {volume} {81}},\ \bibinfo {pages} {938} (\bibinfo {year} {1998})}\BibitemShut
  {NoStop}%
\bibitem [{\citenamefont {Kinoshita}\ \emph {et~al.}(2004)\citenamefont
  {Kinoshita}, \citenamefont {Wenger},\ and\ \citenamefont
  {Weiss}}]{Kinoshita2004}%
  \BibitemOpen
  \bibfield  {author} {\bibinfo {author} {\bibfnamefont {T.}~\bibnamefont
  {Kinoshita}}, \bibinfo {author} {\bibfnamefont {T.}~\bibnamefont {Wenger}}, \
  and\ \bibinfo {author} {\bibfnamefont {D.~S.}\ \bibnamefont {Weiss}},\ }\href
  {\doibase 10.1126/science.1100700} {\bibfield  {journal} {\bibinfo  {journal}
  {Science}\ }\textbf {\bibinfo {volume} {305}},\ \bibinfo {pages} {1125}
  (\bibinfo {year} {2004})},\ \Eprint
  {http://arxiv.org/abs/http://www.sciencemag.org/content/305/5687/1125.full.pdf}
  {http://www.sciencemag.org/content/305/5687/1125.full.pdf} \BibitemShut
  {NoStop}%
\bibitem [{\citenamefont {Paredes}\ \emph {et~al.}(2004)\citenamefont
  {Paredes}, \citenamefont {Widera}, \citenamefont {Murg}, \citenamefont
  {Mandel}, \citenamefont {F\"olling}, \citenamefont {Cirac}, \citenamefont
  {Shlyapnikov}, \citenamefont {H\"ansch},\ and\ \citenamefont
  {Bloch}}]{Paredes2004}%
  \BibitemOpen
  \bibfield  {author} {\bibinfo {author} {\bibfnamefont {B.}~\bibnamefont
  {Paredes}}, \bibinfo {author} {\bibfnamefont {A.}~\bibnamefont {Widera}},
  \bibinfo {author} {\bibfnamefont {V.}~\bibnamefont {Murg}}, \bibinfo {author}
  {\bibfnamefont {O.}~\bibnamefont {Mandel}}, \bibinfo {author} {\bibfnamefont
  {S.}~\bibnamefont {F\"olling}}, \bibinfo {author} {\bibfnamefont
  {I.}~\bibnamefont {Cirac}}, \bibinfo {author} {\bibfnamefont {G.~V.}\
  \bibnamefont {Shlyapnikov}}, \bibinfo {author} {\bibfnamefont {T.~W.}\
  \bibnamefont {H\"ansch}}, \ and\ \bibinfo {author} {\bibfnamefont
  {I.}~\bibnamefont {Bloch}},\ }\href@noop {} {\bibfield  {journal} {\bibinfo
  {journal} {Nature}\ }\textbf {\bibinfo {volume} {429}},\ \bibinfo {pages}
  {277} (\bibinfo {year} {2004})}\BibitemShut {NoStop}%
\bibitem [{\citenamefont {Ho}\ and\ \citenamefont {Yip}(1999)}]{Ho1999}%
  \BibitemOpen
  \bibfield  {author} {\bibinfo {author} {\bibfnamefont {T.-L.}\ \bibnamefont
  {Ho}}\ and\ \bibinfo {author} {\bibfnamefont {S.}~\bibnamefont {Yip}},\
  }\href {\doibase 10.1103/PhysRevLett.82.247} {\bibfield  {journal} {\bibinfo
  {journal} {Phys. Rev. Lett.}\ }\textbf {\bibinfo {volume} {82}},\ \bibinfo
  {pages} {247} (\bibinfo {year} {1999})}\BibitemShut {NoStop}%
\bibitem [{\citenamefont {Wu}\ \emph {et~al.}(2003)\citenamefont {Wu},
  \citenamefont {Hu},\ and\ \citenamefont {Zhang}}]{Wu2003}%
  \BibitemOpen
  \bibfield  {author} {\bibinfo {author} {\bibfnamefont {C.}~\bibnamefont
  {Wu}}, \bibinfo {author} {\bibfnamefont {J.-P.}\ \bibnamefont {Hu}}, \ and\
  \bibinfo {author} {\bibfnamefont {S.-C.}\ \bibnamefont {Zhang}},\ }\href
  {\doibase 10.1103/PhysRevLett.91.186402} {\bibfield  {journal} {\bibinfo
  {journal} {Phys. Rev. Lett.}\ }\textbf {\bibinfo {volume} {91}},\ \bibinfo
  {pages} {186402} (\bibinfo {year} {2003})}\BibitemShut {NoStop}%
\bibitem [{\citenamefont {Lecheminant}\ \emph {et~al.}(2005)\citenamefont
  {Lecheminant}, \citenamefont {Boulat},\ and\ \citenamefont
  {Azaria}}]{Lecheminant2005}%
  \BibitemOpen
  \bibfield  {author} {\bibinfo {author} {\bibfnamefont {P.}~\bibnamefont
  {Lecheminant}}, \bibinfo {author} {\bibfnamefont {E.}~\bibnamefont {Boulat}},
  \ and\ \bibinfo {author} {\bibfnamefont {P.}~\bibnamefont {Azaria}},\ }\href
  {\doibase 10.1103/PhysRevLett.95.240402} {\bibfield  {journal} {\bibinfo
  {journal} {Phys. Rev. Lett.}\ }\textbf {\bibinfo {volume} {95}},\ \bibinfo
  {pages} {240402} (\bibinfo {year} {2005})}\BibitemShut {NoStop}%
\bibitem [{\citenamefont {Cazalilla}\ \emph {et~al.}(2009)\citenamefont
  {Cazalilla}, \citenamefont {Ho},\ and\ \citenamefont {Ueda}}]{Cazalilla2009}%
  \BibitemOpen
  \bibfield  {author} {\bibinfo {author} {\bibfnamefont {M.~A.}\ \bibnamefont
  {Cazalilla}}, \bibinfo {author} {\bibfnamefont {A.~F.}\ \bibnamefont {Ho}}, \
  and\ \bibinfo {author} {\bibfnamefont {M.}~\bibnamefont {Ueda}},\ }\href@noop
  {} {\bibfield  {journal} {\bibinfo  {journal} {New. J. Phys.}\ }\textbf
  {\bibinfo {volume} {11}},\ \bibinfo {pages} {103033} (\bibinfo {year}
  {2009})}\BibitemShut {NoStop}%
\bibitem [{\citenamefont {Gorshkov}\ \emph {et~al.}(2010)\citenamefont
  {Gorshkov}, \citenamefont {Hermele}, \citenamefont {Gurarie}, \citenamefont
  {Xu}, \citenamefont {Julienne}, \citenamefont {Ye}, \citenamefont {Zoller},
  \citenamefont {Demler}, \citenamefont {Lukin},\ and\ \citenamefont
  {Rey}}]{Gorshkov2010}%
  \BibitemOpen
  \bibfield  {author} {\bibinfo {author} {\bibfnamefont {A.~V.}\ \bibnamefont
  {Gorshkov}}, \bibinfo {author} {\bibfnamefont {M.}~\bibnamefont {Hermele}},
  \bibinfo {author} {\bibfnamefont {V.}~\bibnamefont {Gurarie}}, \bibinfo
  {author} {\bibfnamefont {C.}~\bibnamefont {Xu}}, \bibinfo {author}
  {\bibfnamefont {P.~S.}\ \bibnamefont {Julienne}}, \bibinfo {author}
  {\bibfnamefont {J.}~\bibnamefont {Ye}}, \bibinfo {author} {\bibfnamefont
  {P.}~\bibnamefont {Zoller}}, \bibinfo {author} {\bibfnamefont
  {E.}~\bibnamefont {Demler}}, \bibinfo {author} {\bibfnamefont {M.~D.}\
  \bibnamefont {Lukin}}, \ and\ \bibinfo {author} {\bibfnamefont {A.~M.}\
  \bibnamefont {Rey}},\ }\href {\doibase 10.1038/nphys1535} {\bibfield
  {journal} {\bibinfo  {journal} {Nature Physics}\ }\textbf {\bibinfo {volume}
  {6}},\ \bibinfo {pages} {289} (\bibinfo {year} {2010})}\BibitemShut {NoStop}%
\bibitem [{\citenamefont {Krauser}\ \emph {et~al.}(2012)\citenamefont
  {Krauser}, \citenamefont {Heinze}, \citenamefont {Fl{\"a}schner},
  \citenamefont {G{\"o}tze}, \citenamefont {J{\"u}rgensen}, \citenamefont
  {L{\"u}hmann}, \citenamefont {Becker},\ and\ \citenamefont
  {Sengstock}}]{Krauser2012}%
  \BibitemOpen
  \bibfield  {author} {\bibinfo {author} {\bibfnamefont {J.~S.}\ \bibnamefont
  {Krauser}}, \bibinfo {author} {\bibfnamefont {J.}~\bibnamefont {Heinze}},
  \bibinfo {author} {\bibfnamefont {N.}~\bibnamefont {Fl{\"a}schner}}, \bibinfo
  {author} {\bibfnamefont {S.}~\bibnamefont {G{\"o}tze}}, \bibinfo {author}
  {\bibfnamefont {O.}~\bibnamefont {J{\"u}rgensen}}, \bibinfo {author}
  {\bibfnamefont {D.-S.}\ \bibnamefont {L{\"u}hmann}}, \bibinfo {author}
  {\bibfnamefont {C.}~\bibnamefont {Becker}}, \ and\ \bibinfo {author}
  {\bibfnamefont {K.}~\bibnamefont {Sengstock}},\ }\href {\doibase
  10.1038/nphys2409} {\bibfield  {journal} {\bibinfo  {journal} {Nat. Phys.}\
  }\textbf {\bibinfo {volume} {8}},\ \bibinfo {pages} {813} (\bibinfo {year}
  {2012})}\BibitemShut {NoStop}%
\bibitem [{\citenamefont {Krauser}\ \emph {et~al.}(2014)\citenamefont
  {Krauser}, \citenamefont {Ebling}, \citenamefont {Fl{\"a}schner},
  \citenamefont {Heinze}, \citenamefont {Sengstock}, \citenamefont
  {Lewenstein}, \citenamefont {Eckardt},\ and\ \citenamefont
  {Becker}}]{Krauser2014}%
  \BibitemOpen
  \bibfield  {author} {\bibinfo {author} {\bibfnamefont {J.~S.}\ \bibnamefont
  {Krauser}}, \bibinfo {author} {\bibfnamefont {U.}~\bibnamefont {Ebling}},
  \bibinfo {author} {\bibfnamefont {N.}~\bibnamefont {Fl{\"a}schner}}, \bibinfo
  {author} {\bibfnamefont {J.}~\bibnamefont {Heinze}}, \bibinfo {author}
  {\bibfnamefont {K.}~\bibnamefont {Sengstock}}, \bibinfo {author}
  {\bibfnamefont {M.}~\bibnamefont {Lewenstein}}, \bibinfo {author}
  {\bibfnamefont {A.}~\bibnamefont {Eckardt}}, \ and\ \bibinfo {author}
  {\bibfnamefont {C.}~\bibnamefont {Becker}},\ }\href {\doibase
  10.1126/science.1244059} {\bibfield  {journal} {\bibinfo  {journal}
  {Science}\ }\textbf {\bibinfo {volume} {343}},\ \bibinfo {pages} {157}
  (\bibinfo {year} {2014})}\BibitemShut {NoStop}%
\bibitem [{\citenamefont {Ebling}\ \emph {et~al.}(2014)\citenamefont {Ebling},
  \citenamefont {Krauser}, \citenamefont {Fl\"aschner}, \citenamefont
  {Sengstock}, \citenamefont {Becker}, \citenamefont {Lewenstein},\ and\
  \citenamefont {Eckardt}}]{Ebling2014}%
  \BibitemOpen
  \bibfield  {author} {\bibinfo {author} {\bibfnamefont {U.}~\bibnamefont
  {Ebling}}, \bibinfo {author} {\bibfnamefont {J.~S.}\ \bibnamefont {Krauser}},
  \bibinfo {author} {\bibfnamefont {N.}~\bibnamefont {Fl\"aschner}}, \bibinfo
  {author} {\bibfnamefont {K.}~\bibnamefont {Sengstock}}, \bibinfo {author}
  {\bibfnamefont {C.}~\bibnamefont {Becker}}, \bibinfo {author} {\bibfnamefont
  {M.}~\bibnamefont {Lewenstein}}, \ and\ \bibinfo {author} {\bibfnamefont
  {A.}~\bibnamefont {Eckardt}},\ }\href {\doibase 10.1103/PhysRevX.4.021011}
  {\bibfield  {journal} {\bibinfo  {journal} {Phys. Rev. X}\ }\textbf {\bibinfo
  {volume} {4}},\ \bibinfo {pages} {021011} (\bibinfo {year}
  {2014})}\BibitemShut {NoStop}%
\bibitem [{\citenamefont {Ho}\ and\ \citenamefont {Huang}(2015)}]{Ho2015}%
  \BibitemOpen
  \bibfield  {author} {\bibinfo {author} {\bibfnamefont {T.-L.}\ \bibnamefont
  {Ho}}\ and\ \bibinfo {author} {\bibfnamefont {B.}~\bibnamefont {Huang}},\
  }\href {\doibase 10.1103/PhysRevA.91.043601} {\bibfield  {journal} {\bibinfo
  {journal} {Phys. Rev. A}\ }\textbf {\bibinfo {volume} {91}},\ \bibinfo
  {pages} {043601} (\bibinfo {year} {2015})}\BibitemShut {NoStop}%
\bibitem [{\citenamefont {McGuirk}\ \emph {et~al.}(2002)\citenamefont
  {McGuirk}, \citenamefont {Lewandowski}, \citenamefont {Harber}, \citenamefont
  {Nikuni}, \citenamefont {Williams},\ and\ \citenamefont
  {Cornell}}]{McGuirk2002}%
  \BibitemOpen
  \bibfield  {author} {\bibinfo {author} {\bibfnamefont {J.~M.}\ \bibnamefont
  {McGuirk}}, \bibinfo {author} {\bibfnamefont {H.~J.}\ \bibnamefont
  {Lewandowski}}, \bibinfo {author} {\bibfnamefont {D.~M.}\ \bibnamefont
  {Harber}}, \bibinfo {author} {\bibfnamefont {T.}~\bibnamefont {Nikuni}},
  \bibinfo {author} {\bibfnamefont {J.~E.}\ \bibnamefont {Williams}}, \ and\
  \bibinfo {author} {\bibfnamefont {E.~A.}\ \bibnamefont {Cornell}},\ }\href
  {\doibase 10.1103/PhysRevLett.89.090402} {\bibfield  {journal} {\bibinfo
  {journal} {Phys. Rev. Lett.}\ }\textbf {\bibinfo {volume} {89}},\ \bibinfo
  {pages} {090402} (\bibinfo {year} {2002})}\BibitemShut {NoStop}%
\bibitem [{\citenamefont {Nikuni}\ \emph {et~al.}(2002)\citenamefont {Nikuni},
  \citenamefont {Williams},\ and\ \citenamefont {Clark}}]{Nikuni2002}%
  \BibitemOpen
  \bibfield  {author} {\bibinfo {author} {\bibfnamefont {T.}~\bibnamefont
  {Nikuni}}, \bibinfo {author} {\bibfnamefont {J.~E.}\ \bibnamefont
  {Williams}}, \ and\ \bibinfo {author} {\bibfnamefont {C.~W.}\ \bibnamefont
  {Clark}},\ }\href {\doibase 10.1103/PhysRevA.66.043411} {\bibfield  {journal}
  {\bibinfo  {journal} {Phys. Rev. A}\ }\textbf {\bibinfo {volume} {66}},\
  \bibinfo {pages} {043411} (\bibinfo {year} {2002})}\BibitemShut {NoStop}%
\bibitem [{\citenamefont {Fuchs}\ \emph {et~al.}(2003)\citenamefont {Fuchs},
  \citenamefont {Gangardt},\ and\ \citenamefont {Lalo{\"e}}}]{Fuchs2003}%
  \BibitemOpen
  \bibfield  {author} {\bibinfo {author} {\bibfnamefont {J.~N.}\ \bibnamefont
  {Fuchs}}, \bibinfo {author} {\bibfnamefont {D.~M.}\ \bibnamefont {Gangardt}},
  \ and\ \bibinfo {author} {\bibfnamefont {F.}~\bibnamefont {Lalo{\"e}}},\
  }\href@noop {} {\bibfield  {journal} {\bibinfo  {journal} {Eur. Phys. J. D}\
  }\textbf {\bibinfo {volume} {25}},\ \bibinfo {pages} {57} (\bibinfo {year}
  {2003})}\BibitemShut {NoStop}%
\bibitem [{\citenamefont {Endo}\ and\ \citenamefont {Nikuni}(2008)}]{Endo2008}%
  \BibitemOpen
  \bibfield  {author} {\bibinfo {author} {\bibfnamefont {Y.}~\bibnamefont
  {Endo}}\ and\ \bibinfo {author} {\bibfnamefont {T.}~\bibnamefont {Nikuni}},\
  }\href@noop {} {\bibfield  {journal} {\bibinfo  {journal} {J. Low Temp.
  Phys.}\ }\textbf {\bibinfo {volume} {152}},\ \bibinfo {pages} {21} (\bibinfo
  {year} {2008})}\BibitemShut {NoStop}%
\bibitem [{\citenamefont {Du}\ \emph {et~al.}(2008)\citenamefont {Du},
  \citenamefont {Luo}, \citenamefont {Clancy},\ and\ \citenamefont
  {Thomas}}]{Du2008}%
  \BibitemOpen
  \bibfield  {author} {\bibinfo {author} {\bibfnamefont {X.}~\bibnamefont
  {Du}}, \bibinfo {author} {\bibfnamefont {L.}~\bibnamefont {Luo}}, \bibinfo
  {author} {\bibfnamefont {B.}~\bibnamefont {Clancy}}, \ and\ \bibinfo {author}
  {\bibfnamefont {J.~E.}\ \bibnamefont {Thomas}},\ }\href {\doibase
  10.1103/PhysRevLett.101.150401} {\bibfield  {journal} {\bibinfo  {journal}
  {Phys. Rev. Lett.}\ }\textbf {\bibinfo {volume} {101}},\ \bibinfo {pages}
  {150401} (\bibinfo {year} {2008})}\BibitemShut {NoStop}%
\bibitem [{\citenamefont {Natu}\ and\ \citenamefont
  {Mueller}(2009)}]{Natu2009}%
  \BibitemOpen
  \bibfield  {author} {\bibinfo {author} {\bibfnamefont {S.~S.}\ \bibnamefont
  {Natu}}\ and\ \bibinfo {author} {\bibfnamefont {E.~J.}\ \bibnamefont
  {Mueller}},\ }\href {\doibase 10.1103/PhysRevA.79.051601} {\bibfield
  {journal} {\bibinfo  {journal} {Phys. Rev. A}\ }\textbf {\bibinfo {volume}
  {79}},\ \bibinfo {pages} {051601} (\bibinfo {year} {2009})}\BibitemShut
  {NoStop}%
\bibitem [{\citenamefont {Pi{\'e}chon}\ \emph {et~al.}(2009)\citenamefont
  {Pi{\'e}chon}, \citenamefont {Fuchs},\ and\ \citenamefont
  {Lalo{\"e}}}]{Piechon2009}%
  \BibitemOpen
  \bibfield  {author} {\bibinfo {author} {\bibfnamefont {F.}~\bibnamefont
  {Pi{\'e}chon}}, \bibinfo {author} {\bibfnamefont {J.~N.}\ \bibnamefont
  {Fuchs}}, \ and\ \bibinfo {author} {\bibfnamefont {F.}~\bibnamefont
  {Lalo{\"e}}},\ }\href {\doibase 10.1103/PhysRevLett.102.215301} {\bibfield
  {journal} {\bibinfo  {journal} {Phys. Rev. Lett.}\ }\textbf {\bibinfo
  {volume} {102}},\ \bibinfo {pages} {215301} (\bibinfo {year}
  {2009})}\BibitemShut {NoStop}%
\bibitem [{\citenamefont {Ebling}\ \emph {et~al.}(2011)\citenamefont {Ebling},
  \citenamefont {Eckardt},\ and\ \citenamefont {Lewenstein}}]{Ebling2011}%
  \BibitemOpen
  \bibfield  {author} {\bibinfo {author} {\bibfnamefont {U.}~\bibnamefont
  {Ebling}}, \bibinfo {author} {\bibfnamefont {A.}~\bibnamefont {Eckardt}}, \
  and\ \bibinfo {author} {\bibfnamefont {M.}~\bibnamefont {Lewenstein}},\
  }\href {\doibase 10.1103/PhysRevA.84.063607} {\bibfield  {journal} {\bibinfo
  {journal} {Phys. Rev. A}\ }\textbf {\bibinfo {volume} {84}},\ \bibinfo
  {pages} {063607} (\bibinfo {year} {2011})}\BibitemShut {NoStop}%
\bibitem [{\citenamefont {Heinze}\ \emph {et~al.}(2013)\citenamefont {Heinze},
  \citenamefont {Krauser}, \citenamefont {Fl{\"a}schner}, \citenamefont
  {Sengstock}, \citenamefont {Becker}, \citenamefont {Ebling}, \citenamefont
  {Eckardt},\ and\ \citenamefont {Lewenstein}}]{Heinze2013}%
  \BibitemOpen
  \bibfield  {author} {\bibinfo {author} {\bibfnamefont {J.}~\bibnamefont
  {Heinze}}, \bibinfo {author} {\bibfnamefont {J.~S.}\ \bibnamefont {Krauser}},
  \bibinfo {author} {\bibfnamefont {N.}~\bibnamefont {Fl{\"a}schner}}, \bibinfo
  {author} {\bibfnamefont {K.}~\bibnamefont {Sengstock}}, \bibinfo {author}
  {\bibfnamefont {C.}~\bibnamefont {Becker}}, \bibinfo {author} {\bibfnamefont
  {U.}~\bibnamefont {Ebling}}, \bibinfo {author} {\bibfnamefont
  {A.}~\bibnamefont {Eckardt}}, \ and\ \bibinfo {author} {\bibfnamefont
  {M.}~\bibnamefont {Lewenstein}},\ }\href {\doibase
  10.1103/PhysRevLett.110.250402} {\bibfield  {journal} {\bibinfo  {journal}
  {Phys. Rev. Lett.}\ }\textbf {\bibinfo {volume} {110}},\ \bibinfo {pages}
  {250402} (\bibinfo {year} {2013})}\BibitemShut {NoStop}%
\bibitem [{\citenamefont {Koschorreck}\ \emph {et~al.}(2013)\citenamefont
  {Koschorreck}, \citenamefont {Pertot}, \citenamefont {Vogt},\ and\
  \citenamefont {K{\"o}hl}}]{Koschorrek2013}%
  \BibitemOpen
  \bibfield  {author} {\bibinfo {author} {\bibfnamefont {M.}~\bibnamefont
  {Koschorreck}}, \bibinfo {author} {\bibfnamefont {D.}~\bibnamefont {Pertot}},
  \bibinfo {author} {\bibfnamefont {E.}~\bibnamefont {Vogt}}, \ and\ \bibinfo
  {author} {\bibfnamefont {M.}~\bibnamefont {K{\"o}hl}},\ }\href@noop {}
  {\bibfield  {journal} {\bibinfo  {journal} {Nat. Phys.}\ }\textbf {\bibinfo
  {volume} {9}},\ \bibinfo {pages} {405} (\bibinfo {year} {2013})}\BibitemShut
  {NoStop}%
\bibitem [{\citenamefont {Enss}(2015)}]{Enss2015}%
  \BibitemOpen
  \bibfield  {author} {\bibinfo {author} {\bibfnamefont {T.}~\bibnamefont
  {Enss}},\ }\href {\doibase 10.1103/PhysRevA.91.023614} {\bibfield  {journal}
  {\bibinfo  {journal} {Phys. Rev. A}\ }\textbf {\bibinfo {volume} {91}},\
  \bibinfo {pages} {023614} (\bibinfo {year} {2015})}\BibitemShut {NoStop}%
\bibitem [{\citenamefont {Trotzky}\ \emph {et~al.}(2015)\citenamefont
  {Trotzky}, \citenamefont {Beattie}, \citenamefont {Luciuk}, \citenamefont
  {Smale}, \citenamefont {Bardon}, \citenamefont {Enss}, \citenamefont
  {Taylor}, \citenamefont {Zhang},\ and\ \citenamefont
  {Thywissen}}]{Trotzky2015}%
  \BibitemOpen
  \bibfield  {author} {\bibinfo {author} {\bibfnamefont {S.}~\bibnamefont
  {Trotzky}}, \bibinfo {author} {\bibfnamefont {S.}~\bibnamefont {Beattie}},
  \bibinfo {author} {\bibfnamefont {C.}~\bibnamefont {Luciuk}}, \bibinfo
  {author} {\bibfnamefont {S.}~\bibnamefont {Smale}}, \bibinfo {author}
  {\bibfnamefont {A.~B.}\ \bibnamefont {Bardon}}, \bibinfo {author}
  {\bibfnamefont {T.}~\bibnamefont {Enss}}, \bibinfo {author} {\bibfnamefont
  {E.}~\bibnamefont {Taylor}}, \bibinfo {author} {\bibfnamefont
  {S.}~\bibnamefont {Zhang}}, \ and\ \bibinfo {author} {\bibfnamefont {J.~H.}\
  \bibnamefont {Thywissen}},\ }\href {\doibase 10.1103/PhysRevLett.114.015301}
  {\bibfield  {journal} {\bibinfo  {journal} {Phys. Rev. Lett.}\ }\textbf
  {\bibinfo {volume} {114}},\ \bibinfo {pages} {015301} (\bibinfo {year}
  {2015})}\BibitemShut {NoStop}%
\bibitem [{\citenamefont {Koller}\ \emph {et~al.}(2016)\citenamefont {Koller},
  \citenamefont {Wall}, \citenamefont {Mundinger},\ and\ \citenamefont
  {Rey}}]{Koller2016}%
  \BibitemOpen
  \bibfield  {author} {\bibinfo {author} {\bibfnamefont {A.~P.}\ \bibnamefont
  {Koller}}, \bibinfo {author} {\bibfnamefont {M.~L.}\ \bibnamefont {Wall}},
  \bibinfo {author} {\bibfnamefont {J.}~\bibnamefont {Mundinger}}, \ and\
  \bibinfo {author} {\bibfnamefont {A.~M.}\ \bibnamefont {Rey}},\ }\href
  {\doibase 10.1103/PhysRevLett.117.195302} {\bibfield  {journal} {\bibinfo
  {journal} {Phys. Rev. Lett.}\ }\textbf {\bibinfo {volume} {117}},\ \bibinfo
  {pages} {195302} (\bibinfo {year} {2016})}\BibitemShut {NoStop}%
\bibitem [{\citenamefont {Pechkis}\ \emph {et~al.}(2013)\citenamefont
  {Pechkis}, \citenamefont {Wrubel}, \citenamefont {Schwettmann}, \citenamefont
  {Griffin}, \citenamefont {Barnett}, \citenamefont {Tiesinga},\ and\
  \citenamefont {Lett}}]{Pechkis2013}%
  \BibitemOpen
  \bibfield  {author} {\bibinfo {author} {\bibfnamefont {H.~K.}\ \bibnamefont
  {Pechkis}}, \bibinfo {author} {\bibfnamefont {J.~P.}\ \bibnamefont {Wrubel}},
  \bibinfo {author} {\bibfnamefont {A.}~\bibnamefont {Schwettmann}}, \bibinfo
  {author} {\bibfnamefont {P.~F.}\ \bibnamefont {Griffin}}, \bibinfo {author}
  {\bibfnamefont {R.}~\bibnamefont {Barnett}}, \bibinfo {author} {\bibfnamefont
  {E.}~\bibnamefont {Tiesinga}}, \ and\ \bibinfo {author} {\bibfnamefont
  {P.~D.}\ \bibnamefont {Lett}},\ }\href {\doibase
  10.1103/PhysRevLett.111.025301} {\bibfield  {journal} {\bibinfo  {journal}
  {Phys. Rev. Lett.}\ }\textbf {\bibinfo {volume} {111}},\ \bibinfo {pages}
  {025301} (\bibinfo {year} {2013})}\BibitemShut {NoStop}%
\bibitem [{\citenamefont {Giamarchi}(2004)}]{Giamarchi}%
  \BibitemOpen
  \bibfield  {author} {\bibinfo {author} {\bibfnamefont {T.}~\bibnamefont
  {Giamarchi}},\ }\href@noop {} {\emph {\bibinfo {title} {{Quantum Physics in
  one Dimension}}}}\ (\bibinfo  {publisher} {Oxford University Press},\
  \bibinfo {address} {Oxford},\ \bibinfo {year} {2004})\BibitemShut {NoStop}%
\bibitem [{\citenamefont {Jeon}\ and\ \citenamefont {Mullin}(1989)}]{Jeon1989}%
  \BibitemOpen
  \bibfield  {author} {\bibinfo {author} {\bibfnamefont {J.~W.}\ \bibnamefont
  {Jeon}}\ and\ \bibinfo {author} {\bibfnamefont {W.~J.}\ \bibnamefont
  {Mullin}},\ }\href {\doibase 10.1103/PhysRevLett.62.2691} {\bibfield
  {journal} {\bibinfo  {journal} {Phys. Rev. Lett.}\ }\textbf {\bibinfo
  {volume} {62}},\ \bibinfo {pages} {2691} (\bibinfo {year}
  {1989})}\BibitemShut {NoStop}%
\bibitem [{\citenamefont {Boercker}\ and\ \citenamefont
  {Dufty}(1979)}]{Boercker1979}%
  \BibitemOpen
  \bibfield  {author} {\bibinfo {author} {\bibfnamefont {D.~B.}\ \bibnamefont
  {Boercker}}\ and\ \bibinfo {author} {\bibfnamefont {J.~W.}\ \bibnamefont
  {Dufty}},\ }\href {\doibase http://dx.doi.org/10.1016/0003-4916(79)90247-1}
  {\bibfield  {journal} {\bibinfo  {journal} {Annals of Physics}\ }\textbf
  {\bibinfo {volume} {119}},\ \bibinfo {pages} {43 } (\bibinfo {year}
  {1979})}\BibitemShut {NoStop}%
\bibitem [{\citenamefont {Lhuillier}\ and\ \citenamefont
  {Lalo{\"e}}(1982{\natexlab{a}})}]{Lhuillier1982I}%
  \BibitemOpen
  \bibfield  {author} {\bibinfo {author} {\bibfnamefont {C.}~\bibnamefont
  {Lhuillier}}\ and\ \bibinfo {author} {\bibfnamefont {F.}~\bibnamefont
  {Lalo{\"e}}},\ }\href@noop {} {\bibfield  {journal} {\bibinfo  {journal} {J.
  Phys. (Paris)}\ }\textbf {\bibinfo {volume} {43}},\ \bibinfo {pages} {197}
  (\bibinfo {year} {1982}{\natexlab{a}})}\BibitemShut {NoStop}%
\bibitem [{\citenamefont {Lhuillier}\ and\ \citenamefont
  {Lalo{\"e}}(1982{\natexlab{b}})}]{Lhuillier1982II}%
  \BibitemOpen
  \bibfield  {author} {\bibinfo {author} {\bibfnamefont {C.}~\bibnamefont
  {Lhuillier}}\ and\ \bibinfo {author} {\bibfnamefont {F.}~\bibnamefont
  {Lalo{\"e}}},\ }\href@noop {} {\bibfield  {journal} {\bibinfo  {journal} {J.
  Phys. (Paris)}\ }\textbf {\bibinfo {volume} {43}},\ \bibinfo {pages} {225}
  (\bibinfo {year} {1982}{\natexlab{b}})}\BibitemShut {NoStop}%
\bibitem [{\citenamefont {{Jeon, J.W.}}\ and\ \citenamefont {{Mullin,
  W.J.}}(1988)}]{Jeon1988}%
  \BibitemOpen
  \bibfield  {author} {\bibinfo {author} {\bibnamefont {{Jeon, J.W.}}}\ and\
  \bibinfo {author} {\bibnamefont {{Mullin, W.J.}}},\ }\href {\doibase
  10.1051/jphys:0198800490100169100} {\bibfield  {journal} {\bibinfo  {journal}
  {J. Phys. France}\ }\textbf {\bibinfo {volume} {49}},\ \bibinfo {pages}
  {1691} (\bibinfo {year} {1988})}\BibitemShut {NoStop}%
\bibitem [{\citenamefont {Mullin}\ and\ \citenamefont
  {Jeon}(1992)}]{Mullin1992}%
  \BibitemOpen
  \bibfield  {author} {\bibinfo {author} {\bibfnamefont {W.~J.}\ \bibnamefont
  {Mullin}}\ and\ \bibinfo {author} {\bibfnamefont {J.~W.}\ \bibnamefont
  {Jeon}},\ }\href {\doibase 10.1007/BF00126604} {\bibfield  {journal}
  {\bibinfo  {journal} {Journal of Low Temperature Physics}\ }\textbf {\bibinfo
  {volume} {88}},\ \bibinfo {pages} {433} (\bibinfo {year} {1992})}\BibitemShut
  {NoStop}%
\bibitem [{\citenamefont {Lalo{\"e}}\ and\ \citenamefont
  {Mullin}(1990)}]{Laloe1990}%
  \BibitemOpen
  \bibfield  {author} {\bibinfo {author} {\bibfnamefont {F.}~\bibnamefont
  {Lalo{\"e}}}\ and\ \bibinfo {author} {\bibfnamefont {W.}~\bibnamefont
  {Mullin}},\ }\href {\doibase 10.1007/BF01025848} {\bibfield  {journal}
  {\bibinfo  {journal} {Journal of Statistical Physics}\ }\textbf {\bibinfo
  {volume} {59}},\ \bibinfo {pages} {725} (\bibinfo {year} {1990})}\BibitemShut
  {NoStop}%
\bibitem [{\citenamefont {Bogoliubov}(1967)}]{Bogoliubov1967lectures}%
  \BibitemOpen
  \bibfield  {author} {\bibinfo {author} {\bibfnamefont {N.}~\bibnamefont
  {Bogoliubov}},\ }\href {http://books.google.de/books?id=sZsOAAAAQAAJ} {\emph
  {\bibinfo {title} {Lectures on Quantum Statistics}}},\ \bibinfo {series}
  {Lectures on Quantum Statistics}\ No.\ \bibinfo {number} {v. 1}\ (\bibinfo
  {publisher} {Gordon and Breach science publishers},\ \bibinfo {year}
  {1967})\BibitemShut {NoStop}%
\bibitem [{\citenamefont {Snider}(1960)}]{Snider1960}%
  \BibitemOpen
  \bibfield  {author} {\bibinfo {author} {\bibfnamefont {R.~F.}\ \bibnamefont
  {Snider}},\ }\href {\doibase http://dx.doi.org/10.1063/1.1730847} {\bibfield
  {journal} {\bibinfo  {journal} {The Journal of Chemical Physics}\ }\textbf
  {\bibinfo {volume} {32}},\ \bibinfo {pages} {1051} (\bibinfo {year}
  {1960})}\BibitemShut {NoStop}%
\bibitem [{\citenamefont {Thomas}\ and\ \citenamefont
  {Snider}(1970)}]{Thomas1970}%
  \BibitemOpen
  \bibfield  {author} {\bibinfo {author} {\bibfnamefont {M.~W.}\ \bibnamefont
  {Thomas}}\ and\ \bibinfo {author} {\bibfnamefont {R.~F.}\ \bibnamefont
  {Snider}},\ }\href {\doibase 10.1007/BF01009711} {\bibfield  {journal}
  {\bibinfo  {journal} {Journal of Statistical Physics}\ }\textbf {\bibinfo
  {volume} {2}},\ \bibinfo {pages} {61} (\bibinfo {year} {1970})}\BibitemShut
  {NoStop}%
\end{thebibliography}%
\end{document}